\documentclass[11pt, a4paper]{article}
\usepackage{amsmath ,amsthm ,amssymb}
\usepackage{graphicx}
\usepackage{geometry} 
\usepackage{slashed}
\usepackage{braket}
\usepackage[colorlinks = true,
            linkcolor = blue,
            urlcolor  = cyan,
            citecolor = blue,
            anchorcolor = blue]{hyperref}
\usepackage{relsize}
\usepackage{bbm}
\usepackage{mathtools} 
\usepackage[utf8]{inputenc}
\usepackage{stmaryrd} 
\usepackage{tikz-cd}
\usepackage{tikz}
\usetikzlibrary{matrix}
\usepackage{bbold}
\usepackage{cleveref}

\usepackage{xcolor}

\def\susp{\uparrow}

\begin{document}

\begin{titlepage}

\setcounter{page}{1}
\begin{flushright}
LMU-ASC 19/22\\
\end{flushright}
\begin{center}
{\Large \bf Spin fields for the spinning particle}
\vskip 1cm

\vspace{12pt}

E. Boffo\,$^{1}$, I. Sachs\,$^2$\\

\vskip 25pt

{\em $^1$ \hskip -.1truecm  Faculty of Mathematics and Physics, Mathematical Institute, Univerzita Karlova,\\ Sokolovsk\'{a} 49/83, 186 75 Praha, Czech Republic \vskip 5pt }

\vskip 10pt
 {\em $^2$ \hskip -.1truecm Arnold Sommerfeld Center for Theoretical Physics,
Ludwig-Maximilian-Universit\"at ,\\
Theresienstr. 37, D-80333 M\"unchen, Germany}

\end{center}

\vskip 15pt

\begin{center}

{emails: \small{{\tt boffo@karlin.mff.cuni.cz},  
{\tt Ivo.Sachs@physik.uni-muenchen.de}}}
\end{center}

\vskip .5cm

\begin{abstract}
We propose an analogue of spin fields for the relativistic RNS-particle in 4 dimensions, in order to describe Ramond-Ramond states as "two-particle" excitations on the world line. On a natural representation space we identify a differential whose cohomology agrees with RR-fields equations. We then  discuss the non-linear theory encoded in deformations of the latter by background fields. 
We also formulate a sigma model for this spin field from which we recover the RNS-formulation by imposing suitable constraints. 
\end{abstract}

\end{titlepage}

\newpage

\tableofcontents

\date{}

\section{Introduction}
In the RNS formulation of string theory, spin fields appear in the world sheet CFT to relate the Ramond (R) vacuum to the Neveu-Schwarz (NS) vacuum. While these fields do not appear in the string sigma model, such operators are necessary to represent vertices involving R-fields and NS-fields, such as the coupling of quarks and gluons. In addition, spin fields are indispensable in the RNS formulation to construct vertex operators of RR states.  
The construction of spin fields arises rather naturally within the bosonization of the RNS world sheet fermion \cite{Friedan:1985ge}. Geometrically, they produce a cut in the world-sheet since their OPE is not single valued. On the other hand, introducing off-shell RR fields as backgrounds in the RNS world-sheet description of string theory has not been successful, since the corresponding vertex operators do not have a familiar geometric interpretation. This is in contrast to massless NS or NS-NS background fields which in a suitable picture arise rather naturally as a deformation of the string world sheet theory. An alternative route to couple RR-fields in string theory is to use the Green-Schwarz formulation \cite{Green:1983wt} through the pure spinor aproach \cite{Berkovits:2002zk}. One may also  resort to covariant super string field theory \cite{Berkovits:1995ab,Sen:2014dqa,Erler:2015lya}. This approach, while conceptually simple and intuitive is perturbative in nature. Still, this framework is useful to discuss off-shell aspects such as mass renormalization \cite{Pius:2014iaa} or deformations of D-branes \cite{Mattiello:2019gxc,Maccaferri:2019ogq,Vosmera:2019mzw} among other. See also \cite{Cho:2018nfn,Sen:2020cef} for recent applications of superstring field theory to RR-backgrounds.  

To explore such issues, the world line, as a toy model of string theory, on one side and as an alternative construction of ordinary quantum field theory on the other, has often come in useful. For instance, the Brink-Schwarz particle \cite{Brink:1981nb} has been instrumental in the development of the pure spinor formulation of string theory \cite{Berkovits:2002zk}. From the latter, the low energy effective field equations for the NS-sector can be obtained in a rather elegant manner by imposing nilpotency of the pure spinor BRST operator \cite{Berkovits:2001ue}.  Similarly, the RNS world line has been shown to reproduce the low energy effective action for the NS-NS sector of string theory by demanding that the BRST operator of the world line theory squares to zero on a suitable representation space \cite{Dai:2008bh,Bonezzi:2018box,Bonezzi:2020jjq}. This is tantamount to the condition that massless perturbative excitations can propagate consistently on this background. On the world line the construction of the BRST-operator is manifestly background independent, a luxury we do not have in string theory. In particular we can use the world line to explore non-perturbative aspects, such as backgrounds with no metric \cite{Bockisch:2022eas} which are hard to explore even in ordinary Einstein gravity. 

Given this state of affairs one might hope that the world line could also play a role in describing the coupling of RR-background fields once Ramond states can be successfully implemented in it. This is the purpose of this note. Since bosonization is not available on the world line and neither is the operator product expansion it is not, a priory, clear how to construct the analog of spin fields on the world line.  What does carry over from the world sheet to the world line, are equal time commutators, via contour integrals, which contains much less information, making it harder to guess what the "resolution" of the world line fermion should be. 

Taking inspiration form \cite{Sorokin:1988nj} (see also \cite{Volkov:1988vf}), in section \ref{sec:resolution}, we propose to mimic the world sheet bosonization by introducing a pair of Grassmann even space-time spinors $\vartheta^\alpha$ and $\lambda_\alpha$ together with a Grassmann degree shifting operator $\susp$. Recall that, in contrast to two dimensions, fractional  statistics is not available in one dimension due to the absence of branch cuts. To fill this gap 
we introduce the Grassmann parity shifting operator $\susp$.  

We then show, in section \ref{sec:N=1_BRST}, that on a suitable representation space this resolution is equivalent to the $\mathcal{N}=1$ supersymmetric spinning world line, which is the point particle limit of the Ramond sector. In particular, it leads to an equivalent BRST-complex. One advantage of resolving the world line spinor is that it allows to identify a sub-complex of the BRST complex of the  $\mathcal{N}=1$ world line whose differential $\mathbf{q}$ is the chiral half of the 
$\mathcal{N}=1$ supercharge, in particular, eliminating the need for a $(b,c)$ system. Nilpotency then reflects the decoupling of off-shell anti-chiral massless Ramond states. Furthermore, deformations of the differential by background fields (unlike the standard BRST charge $Q$) leads to the correct coupling of Yang-Mills potentials to world line Feynman graphs following the standard construction \cite{Dai:2008bh}.  

After this preparation, we consider the RR-sector in section \ref{Multiparticle states} and show that the cohomology agrees with expectations for the RR-field strengths in four dimensions. Nilpotency of the world line supercharge now reflects the decoupling of off-shell RR-states. Infinitesimal deformations of $\mathbf{q}$ by background fields are related to various RR-states by action on a reference state $\ket{\Omega}$ in its module, akin of the operator-state correspondence in string theory. The deformed charge fails to be nilpotent for generic backgrounds. For some deformations this is compensated by elimination of the previously exact states from the perturbative spectrum. For the remaining ones we show that there is a consistent truncation setting them to zero, akin of the GSO-projection in type II strings. 
Finally, we propose a world line sigma model for the spin field in section \ref{sigma_model}, from which the resolved $\mathcal{N}=1$ spinning particle  is recovered as a constraint system of the BRST world line action of the former.

We close by describing the relation between this sigma model and the Brink-Schwarz particle. The conventions that we adopted are listed in the Appendix.

\section{Resolution of the RNS world line fermion}\label{sec:resolution}
Let us begin by reviewing the standard BRST quantization of the spinning world line \cite{Brink:1976sc} before introducing the analogue of spin fields for the latter. 
\subsection{Standard BRST Quantization of the world line}
Upon gauge fixing the world line action 
\begin{equation}
S=\int \mathrm{d}\tau \big[p_\mu\dot x^\mu+\mathrm{i}\psi_\mu\dot\psi^\mu-\tfrac{e}{2}\,p^2-\mathrm{i}\chi\,\psi\!\cdot p\big]\;,    
\end{equation}
with $\mathcal{N}=1$ world line SUSY, one constructs the BRST operator 
\begin{align}\label{BRSTN10}
    Q_0=-cH_0+\gamma q_0 +b\gamma^2\,,
\end{align}
where $q_0=\psi^\mu p_\mu$ is the world line supercharge with associated super-reparametrization ghosts $\gamma$ and $\beta$, and $H_0$ is the world line Hamiltonian, with associated reparametrization ghost-antighost pair $(c,b)$. In accordance with the zero mode of the world sheet fermion the action of $\psi^\mu$ on one-particle spinors, $\ket{\phi}=\phi_{a}\ket{a}$ in space-time is taken to be that of the gamma matrices, that is 
\begin{align}
    \psi^\mu\;\phi_{a}\ket{a}=(\Gamma^\mu)_{a'a}\phi_a\ket{a'}\,.
\end{align}
Then, $Q_0\ket{\phi}=0$ reproduces the Dirac equation, $-\mathrm{i}(\partial
\!\!\!/\phi)_{a}=0$. 

For the action with $\mathcal{N}=2$ world line SUSY
\begin{equation}
S=\int \mathrm{d}\tau \big[p_\mu\dot x^\mu+\mathrm{i}\bar \psi_\mu\dot\psi^\mu-\tfrac{e}{2}\,p^2-\mathrm{i}\chi\,\bar\psi\!\cdot p-\mathrm{i}\bar\chi\,\psi\!\cdot p\big]\;,    
\end{equation}
we have instead two supercharges and  correspondingly, complexified superghosts satisfying $[\gamma, \bar\beta] = 1 = [\bar \gamma, \beta]$ (together with $\{c,b\} =1$). From this one deduces the BRST operator
\begin{align}\label{eq:BRSTN20}
    Q_0=cH_0+\bar\gamma q_0+\gamma \bar q_0 +b\gamma\bar\gamma\,.
\end{align}
One particle states are generated by the action of $\psi^\mu,\beta,\gamma$ and $c$ on a ground state $\ket{0}$. The states are thus given by wave functions $\phi(x,\psi^\mu,\beta,\gamma,c)$ which can be taken to be eigenstates of the world line $U(1)_R$-charge, $J$ \cite{Dai:2008bh}. 

What is missing for the $\mathcal{N}=1$ world-line, in comparison to string theory, is the possibility to create a spinor state $\ket{a}$ out of the vacuum $\ket{0}$. This is necessary, for example, in order to describe interactions of fermionic internal lines with other fields in world lines Feynman amplitudes. In string theory space-time fermions are created by acting  with spin fields on the vacuum. The latter are, however, not readily available on the world line since there is no \emph{bosonization} of world line fermions.\footnote{Mathematically, bosonization in fact consists of forming complex linear combinations of the Clifford algebra $Cl(2n)$ generators, that transform in the fundamental of $SU(n)$.}
To fill this gap we will now propose an alternative representation of $Q_0$ in \eqref{BRSTN10} by resolving $\psi^\mu$ in analogy with the spin fields in string theory.

\subsection{Resolution of $\psi^\mu$}
\label{Resolution}
In the canonical approach the world line variables ($\psi^\mu$, $x^\mu$, $p^\mu$) generate a graded 
associative algebra $\mathcal{A}$ endowed with a graded Lie bracket,
\begin{align} \label{gLa}
    [x^\mu,p_\nu]=\mathrm{i}\delta^\mu_{\;\nu}\,,\quad \{\psi^\mu,\psi^\nu\}=-2\eta^{\mu\nu}\,.
\end{align}
The idea is to find an alternative expression for $\psi$ in terms of new \emph{bosonic} (Weyl) generators $\vartheta^\alpha, \varepsilon^\alpha, \lambda_\alpha$ 
and a product $\circ$, together with the sole relations,\footnote{To avoid confusion: $\lambda_\alpha$ is not a ghost as in \cite{Berkovits:2001ue}. Rather it is just another world line variable as in \cite{Sorokin:1988jor}. } 
\begin{align}\label{FFCR0}
\left[\lambda_{\alpha}, \vartheta^{\beta}\right]&= \lambda_\alpha \circ \vartheta^\beta - \vartheta^\beta \circ \lambda_\alpha = \delta_{\alpha}^{\beta}\,, \qquad \left[\lambda_\alpha, \varepsilon^\beta\right] = \delta_\alpha^\beta, \qquad [\lambda_\alpha,\lambda_\beta]= 0\,.
\end{align}
In particular, the above commutator is not going to make up a Lie bracket, because, for instance, we will violate Jacobi identity
\begin{align}\label{eq:no_jac}
    \left[\left[\lambda_{\alpha}, \vartheta^{\beta}\right],\vartheta^\gamma\right] + \left[\vartheta^\beta, \left[\lambda_\alpha,\vartheta^\gamma\right]\right] =\left[\delta_\alpha^\beta, \vartheta^\gamma\right] + \left[\vartheta^\beta, \delta_\alpha^\gamma\right] = 0 \neq \left[\lambda_\alpha, \left[\vartheta^\beta,\vartheta^\gamma\right]\right],
\end{align}
since we have not imposed  $\left[\vartheta^\alpha,\vartheta^\beta\right]=0$ at this point.\footnote{While it would be consistent to impose $\left[\vartheta^\alpha,\vartheta^\beta\right]=0$ at the level of the algebra, this would cause problems when representing this algebra on a vector space in section \ref{Multiparticle states}.} In view of comparing our construction to spin fields it is helpful to recall that in a world sheet theory  commutators are well defined only for operators whose OPE does not have branch cuts, which is not the case for spin fields in generic space-time dimensions. This feature is reflected in \cref{eq:no_jac} for a 1-dimensional source space. 

Based on these two observations we declare the extended set $(x^\mu, p_\nu, \vartheta^\alpha, \varepsilon^\beta, \lambda_\gamma, \mathbb{1})$ to form just a vector space endowed with the relations
\begin{align}\label{gLa_no_psi}
    [x^\mu,p_\nu]=\mathrm{i}\delta^\mu_{\;\nu}\,,\quad \left[\lambda_{\alpha}, \vartheta^{\beta}\right]&= \delta_{\alpha}^{\beta} = \left[\lambda_\alpha, \varepsilon^\beta \right]\,. 
\end{align}
In order to construct a real space-time vector we need to add a second set of elements which we denote by $\tilde{\vartheta}_{\dot{\beta}}, \tilde\varepsilon_{\dot\beta}$ and $\tilde{\lambda}^{\dot{\alpha}}$, together with an extension of the product $\circ$ so that
\begin{align}\label{FFCR1}
\left[\tilde{\lambda}^{\dot{\alpha}},\tilde{\vartheta}_{\dot{\beta}} \right]= \tilde\lambda^{\dot\alpha} \circ \tilde{\vartheta}_{\dot\beta} - \tilde{\vartheta}_{\dot\beta} \circ \tilde\lambda^{\dot\alpha} = \delta_{\dot{\alpha}}^{\dot{\beta}}, \qquad \left[\tilde\lambda^{\dot\alpha} , \tilde\varepsilon_{\dot\beta}\right] = \delta^{\dot\alpha}_{\dot\beta} \,.
\end{align}
The second copy $\varepsilon^\alpha$ of Weyl spinors introduced above will not play any role at the moment but will be convenient to describe Ramond-Ramond states in section \ref{Multiparticle states}. 

We will now argue that we can recover the Clifford algebra \eqref{gLa} by introducing, in addition, a Grassmann degree-shifting involution $\susp$, $\susp \susp = 1$, with  the identification\footnote{After absorbing $\susp$ in $\vartheta$, a classically (but not canonically) equivalent expression can be found in \cite{Volkov:1988vf}.}  
\begin{align}
    \psi^{\mu} :=  \left(\vartheta^{\alpha}\sigma^{\mu}_{\alpha \dot{\alpha}}  \tilde{\lambda}^{\dot{\alpha}} +\tilde{\vartheta}_{\dot{\beta}}\tilde{\sigma}^{\mu \; \dot{\beta} \beta} \lambda_{\beta}\right) \susp \,. \label{first-gamma}
\end{align}
Here $ \sigma^\mu$ is the extended set of $2\times2$ dimensional Pauli matrices, $\sigma^\mu := \begin{pmatrix} \text{id}, \sigma^i \end{pmatrix}$, while in $\tilde{\sigma}^\mu$ the spatial part comes with the opposite sign. Our conventions are listed in the Appendix. The degree-shifting operator $\susp$ is needed for $\psi^\mu$ to have the correct Grassmann parity. The task of finding a matrix representation for $\susp$ will not be addressed because it is not essential for what follows: such operator just serves the purpose of implementing the analogue of a branch cut as a $\mathbb{Z}_2$ map on the world line.
Notice that reality of $\psi^\mu$ requires that $\vartheta^{\beta}$ and $\tilde\vartheta_{\dot{\beta}}$ form a Majorana spinor. This is just one of the possibilities at our disposal in order to build a real $\psi^\mu$, and equivalent results stem from using $\varepsilon$ and $\tilde\varepsilon$. The key property used in the quantization of the spinning particle is that the world line fermion $\psi^{\mu}$ acts on a state $ \varphi_\alpha\ket{\alpha}+\chi^{\dot\alpha} \ket{\dot\alpha}$ of the  quantized theory as a Gamma matrix. More concretely, we choose the convention that the Ramond 1-particle states $\ket{\alpha}$ and $\ket{\dot\alpha}$ are created by the \emph{"spin fields"} $\vartheta^\alpha$ or $\tilde{\vartheta}_{\dot\alpha}$ on a ground state $\ket{0}$, represented by the constant function, or alternatively, on that of shifted parity $\susp \ket{0}$. In contrast, $\varepsilon$ and $\tilde\varepsilon$ alone cannot be used as creators since they do not lead to a $\psi^\mu$ invariant subspace. Then from \eqref{gLa_no_psi} we read off that $\lambda_\alpha$ and $\tilde\lambda^{\dot\alpha}$ act as annihilators operators. Thus
\begin{equation}\label{eq:kets}
\begin{matrix}
    \ket{\alpha}=\vartheta^\alpha \ket{0}\,, &  \ket{\dot\alpha}=\tilde\vartheta_{\dot\alpha} \ket{0}\,,\\
    \ket{\alpha}^\susp = \vartheta^\alpha \susp \ket{0}\, , & \ket{\dot\alpha}^\susp = \vartheta_{\dot\alpha}\susp \ket{0} \, ,
\end{matrix}
\end{equation}
are linear functions in $\vartheta$ or $\tilde{\vartheta}$. The space generated by \eqref{eq:kets} consists of two sectors, of even or odd Grassmann parity (first row vs. second row), mapped into each other by $\psi^\mu$. They give rise to wave functions
\begin{equation}\label{eq:Psi_wav}
\ket{\Psi}=\phi_\alpha \ket{\alpha} + \phi^{\susp}_\alpha \ket{\alpha}^\susp + \tilde{\chi}^{\dot\alpha} \ket{\dot\alpha} + \tilde{\chi}^{\susp\dot{\alpha}} \ket{\dot\alpha}^\susp.
\end{equation}
One of the benefits of resolving $\psi^\mu$ in terms of $\lambda_{\alpha}$ and $\vartheta^{\beta}$ is thus clear: in analogy with the action of spin fields in string theory, the Ramond states $\ket{\alpha}$ and $\ket{\dot \alpha}$ can be obtained by either acting with $\vartheta^{\beta}$ and $\tilde\vartheta_{\dot\alpha}$ on the invariant ground state $\ket{0}$. If we assume the graded commutativity,
\[ \lambda_\alpha \susp = \susp \lambda_\alpha \, , \qquad \tilde\lambda^{\dot\alpha} \susp = \susp \tilde{\lambda}^{\dot\alpha}\, ,\]
in the Hilbert space $\mathcal{H}$ spanned by the previously described states \eqref{eq:kets}, the world line fermion \eqref{first-gamma} realizes the Clifford algebra 
\begin{align}\label{eq:pmpn}
    \{\psi^\mu,\psi^\nu\}\vert_\mathcal{H}&= \{\vartheta^{\alpha}\sigma^{\mu}_{\alpha\dot\alpha}\tilde{\lambda}^{\dot{\alpha}} \susp, \tilde{\vartheta}_{\dot{\beta}}\tilde{\sigma}^{\nu \; \dot{\beta} \beta} \lambda_{\beta} \susp\}\vert_{\mathcal{H}} + \{\tilde{\vartheta}_{\dot{\beta}} \tilde{\sigma}^{\mu \; \dot{\beta} \beta} \lambda_{\beta} \susp, \vartheta^\alpha \sigma^{\nu}_{\alpha \dot{\alpha}} \tilde{\lambda}^{\dot{\alpha}}\susp\}\vert_\mathcal{H} \nonumber \\
    &=\big(\big(\sigma^{\mu}_{\alpha\dot\alpha}\tilde{\sigma}^{\nu \; \dot{\alpha} \beta} + \sigma^{\nu}_{\alpha \dot{\alpha}} \tilde{\sigma}^{\mu \; \dot{\alpha} \beta} \big)\vartheta^{\alpha}\lambda_{\beta}+\big(\tilde{\sigma}^{\nu \; \dot{\beta} \alpha} \sigma^{\mu}_{\alpha\dot\alpha} + \tilde{\sigma}^{\mu \; \dot{\beta} \alpha} \sigma^{\nu}_{\alpha\dot\alpha}\big)\tilde{\vartheta}_{\dot{\beta}} \tilde{\lambda}^{\dot{\alpha}}\big)\vert_\mathcal{H} \nonumber\\
    &=-2\eta^{\mu\nu}\big(\vartheta^{\beta}\lambda_{\beta}+\tilde{\vartheta}_{\dot{\beta}}\tilde{\lambda}^{\dot{\beta}}\big)\vert_\mathcal{H}.
\end{align}
The minus sign is coherent with our choice of signature of the Minkowski metric $\eta^{\mu \nu}$, taken to be mostly positive. Defining $\mathcal{I} := \vartheta^\alpha \lambda_\alpha + \tilde\vartheta_{\dot\alpha}\tilde\lambda^{\dot\alpha}$, one can easily observe that $\mathcal{H}$ is contained in the kernel of $\mathcal{I}-1$ and therefore $\psi^\mu$ does indeed realize the Clifford algebra and $\mathcal{H}$ its module
\begin{align}
    \left\{\psi^{\mu}, \psi^{\nu}\right\} \vert_{\mathcal{H}} = -2\eta^{\mu \nu}\,.
\end{align}

We will also need 
\begin{align}
    [\psi^\mu,\psi^\nu]\vert_\mathcal{H}& 
     =\big(-4\mathrm{i} (\sigma^{\mu \nu})_{\alpha}{}^{\beta} \vartheta^\alpha \lambda_{\beta} - 4 \mathrm{i} (\tilde{\sigma}^{\mu \nu})^{\dot{\alpha}}{}_{\dot{\beta}} \tilde{\vartheta}_{\dot{\alpha}} \tilde{\lambda}^{\dot{\beta}}  \nonumber\\ & \quad + \vartheta^\alpha \susp \tilde{\vartheta}_{\dot{\beta}} \susp \big(\sigma^{\mu}{}_{\alpha \dot{\alpha}} \tilde{\sigma}^{\nu \; \dot{\beta}\beta} - \sigma^{\nu}{}_{\alpha \dot{\alpha}} \tilde{\sigma}^{\mu \; \dot{\beta}\beta} \big) \tilde{\lambda}^{\dot{\alpha}} \lambda_{\beta}  \nonumber \\ & \quad + \vartheta^\alpha \susp \tilde{\vartheta}_{\dot{\beta}} \susp \big(\tilde{\sigma}^{\nu \; \dot{\beta} \beta} \sigma^{\mu}{}_{ \alpha \dot{\alpha}} - \tilde{\sigma}^{\mu \; \dot{\beta} \beta} \sigma^{\nu}{}_{ \alpha \dot{\alpha}} \big) \tilde{\lambda}^{\dot{\alpha}} \lambda_\beta \nonumber \\ & \quad -2\mathrm{i}(\tilde\sigma^{\mu\nu})_{\dot{\alpha}\dot{\beta}}\tilde{\lambda}^{\dot{\alpha}}\tilde{\lambda}^{\dot{\beta}}\vartheta^\gamma \susp \vartheta_\gamma \susp  -2\mathrm{i}(\sigma^{\mu\nu})_{{\alpha}{\beta}}{\lambda}^{{\alpha}}{\lambda}^{{\beta}}\tilde\vartheta_{\dot\gamma} \susp \tilde\vartheta^{\dot\gamma} \susp\big)\vert_\mathcal{H} \label{SOalg}
\end{align}
On the module generated by \eqref{eq:kets} the last three lines vanish and thus 
\begin{align}
    -[\psi^\mu,\psi^\nu]\vert_\mathcal{H} &=4\mathrm{i} (\sigma^{\mu\nu})_\alpha^{\;\;\beta}\vartheta^{\alpha}\lambda_{\beta} +4\mathrm{i} (\tilde\sigma^{\mu\nu})^{\dot\beta}_{\;\;\dot\alpha}\tilde{\vartheta}_{\dot{\beta}}\tilde{\lambda}^{\dot{\alpha}}  \,,
\end{align}  
which is consistent with expectations. The minus sign follows from our conventions. We could have arrived at the same observation by recalling the canonical embedding of the $\mathfrak{so}(1,3)$ algebra in the Clifford algebra. 

While we will not specifically rely on it for vast part of this article, it is possible to define an inner product on $\mathcal{H}$ such that $\psi^\mu$ is self-adjoint with respect to it. Introducing the variables $\theta^\alpha=\vartheta^\alpha\susp$ and $\tilde\theta_{\dot\alpha}=\tilde\vartheta_{\dot\alpha}\susp$ such an inner product can be defined as 
\begin{align}
    (\ket{\Psi},\ket{\Phi})=&\int d\theta_1 d\theta_2 \; \psi^{\susp *}_\alpha(x)\theta^\alpha  \varphi^\susp_\beta(x)\theta^\beta  -\int d\tilde\theta_1 d\tilde\theta_2 \; \tilde\psi^{\susp *\dot{\alpha}}(x)\tilde\theta_{\dot{\alpha}}\tilde\varphi^{\susp\dot{\alpha}}(x)\tilde\theta_{\dot{\alpha}}\nonumber\\&+\int d\theta_1 d\theta_2 \; \psi^*_\alpha(x)\theta^\alpha  \varphi_\beta(x)\theta^\beta  -\int d\tilde\theta_1 d\tilde\theta_2 \; \tilde\psi^{*\dot{\alpha}}(x)\tilde\theta_{\dot{\alpha}}\tilde\varphi^{\dot{\alpha}}(x)\tilde\theta_{\dot{\alpha}}\,. \label{hermitian-metric}
\end{align}

 \section{Ramond sector}\label{sec:N=1_BRST}

Let us now consider the dynamical system defined by the BRST operator \eqref{BRSTN10} with $\psi^\mu$ expressed in terms of $\vartheta, \lambda$ and $\susp$ and explore its properties. As in the previous section, we take the module $\mathcal{H}$ to be the $\psi^\mu$-invariant kernel of $\mathcal{I}-1$ spanned by \eqref{eq:kets}. In accordance with the Ramond sector of string theory, we define the vacuum state in the ghost sector by $\beta{\ket{0}}=0 = \beta \susp \ket{0}$, while $\gamma{\ket{0}}\neq 0 \neq \gamma \susp \ket{0}$. Then $\mathcal{H}$ contains generic states of the form\footnote{This corresponds to the polarisation where $p^\mu$, $b$ and $\beta$ acts as derivative operators so that a generic wave function is of the form $\Phi(x,\vartheta^\beta,\tilde\vartheta_{\dot{\alpha}},c,\gamma)$. } 
\begin{align} \label{Rfield}
    \ket{\Phi}=&\varphi_\alpha(x) \vartheta^{\alpha}+\tilde\chi^{\dot{\alpha}}(x)\tilde\vartheta_{\dot{\alpha}}+\varphi^\susp_\alpha(x)\vartheta^\alpha \susp  +\tilde\chi^{\susp\dot{\alpha}}(x) \tilde{\vartheta}_{\dot{\alpha}}\susp+\cdots\nonumber\\ 
    &+\gamma^n\varphi^{(n)}_\alpha(x)\vartheta^\alpha+\gamma^n\varphi^{\susp(n)}_\alpha(x) \vartheta^{\alpha}\susp+\cdots +c\gamma^n\varphi^{c(n)}_\alpha(x)\vartheta^\alpha+\cdots
\end{align}
where $\cdots$ stands for further terms with arbitrary powers of $\gamma$ and $c$ (the $c$-dependence is of course at most linear). Using \eqref{eq:pmpn}, we first find from nilpotency, that $H=-2\mathcal{I} \eta^{\mu\nu}p_\mu p_\nu$, where $\mathcal{I}=(\vartheta^{\beta}\lambda_{\beta}+\tilde{\vartheta}_{\dot{\beta}}\tilde{\lambda}^{\dot{\beta}})$ acts as the identity operator on $\mathcal{H}$.  Furthermore, $Q_0$ preserves $\mathcal{H}$.

At ghost number 0 the cohomology of $Q_0$ at odd Grassmann parity is generated by wave functions of the form  
\begin{align}\label{eq:ferm}
    \ket{\Phi}=\varphi^\susp_\alpha(x) \vartheta^{\alpha} \susp+\tilde\chi^{\susp \,\dot{\alpha}}(x)\tilde\vartheta_{\dot{\alpha}} \susp
\end{align}
with $\partial\!\!\!/\tilde\chi^\susp=\tilde\partial\!\!\!/\varphi^\susp =0$. Thus, we recover the same results as in the standard formulation.

At generic ghost number, the space of fields is infinite dimensional: polynomials of  $c$ and arbitrary order in $\gamma$ form an infinite tower of fields. With the obvious redefinition $\tilde\chi^{\susp c(0) \, \dot\alpha}(x) \equiv \tilde\chi^{\susp c \, \dot\alpha}(x)$, we see that for $(\varphi_\alpha, \varphi^{(1)}_\alpha, \dots, \tilde{\chi}^{\susp c(0) \dot\alpha}, \tilde{\chi}^{\susp c(1) \dot\alpha}, \dots)$ we get a coupled system of equations: 
\begin{subequations}
\begin{align}
c\, \square \varphi_\alpha(x) \vartheta^\alpha &=0  \nonumber\\
- \mathrm{i} \gamma (\tilde\vartheta\sigma^\mu)^\alpha \partial_\mu \varphi_\alpha(x) \susp & = 0 \nonumber\\
    \square \varphi^{(n)}_{\alpha} \vartheta^\alpha - \mathrm{i} \partial_\mu \tilde\chi^{\susp c(n-1) \; \dot\alpha} (\vartheta\sigma^{\mu})_{\dot\alpha}  &=0 \; \; \;  \text{at $c\gamma^{n}$, with $n\geq1$}\label{2ord}\\
    -\mathrm{i} \partial_\mu \varphi^{(n)}_{\alpha}(\tilde{\vartheta}\tilde{\sigma}^{\mu})^{\alpha} + \tilde\chi^{\susp c(n-1) \, \dot\alpha} \tilde\vartheta_{\dot\alpha} &=0  \; \; \text{at $\gamma^{n+1}$, $n\geq 1$}\,. \label{1ord}
\end{align}
\end{subequations}
Identical conditions hold for suitable combinations of the fields in \eqref{Rfield}, with various chiralities and Grassmann parity. Examining the system, it is not difficult to notice that the field in ghost number $0$ does not receive additional contributions from the ghosts. Concerning higher ghost degrees, one can be easily convinced, from equation \eqref{2ord} and \eqref{1ord}, that  $\tilde\chi^{\susp c(n-1)}$ is determined by $\varphi^{(n)}$ which is unconstrained. Note that these properties are not peculiarities of the resolution of $\psi^\mu$ but are already present in the standard BRST quantization of the $\mathcal{N}=1$ particle.

\subsection{Coupling to background fields}
\label{backgrounds}
It is well known that the $\mathcal{N}=1$ world line in the standard formulation can be coupled to a background Yang-Mills field (e.g. \cite{Schubert:2001he,Dai:2008bh}) without inducing any dynamical equations on the latter. This is tantamount to the substitution $p_\mu\to p_\mu+A_\mu  =: \Pi_\mu$ in the supercharge, i.e.
\begin{align}
    q=q_0+\psi^\mu A_\mu= q_0+ \left(\vartheta^{\alpha}A_{\alpha \dot{\alpha}}  \tilde{\lambda}^{\dot{\alpha}} +\tilde{\vartheta}_{\dot{\beta}}\tilde{A}^{\dot{\beta} \beta} \lambda_{\beta}\right) \susp \,, \label{eq:A_q}
\end{align}
which, in turn, induces the non-minimal coupling 
\begin{align}\label{eq:N=1_back}
    H= \{\psi^\mu, \psi^\nu\} \Pi_\mu \Pi_\nu+[\Pi_\mu ,\Pi_\nu]\psi^\mu\psi^\nu
\end{align}
in the Hamiltonian. The commutator $[H,q]$, where $q:=q_0 + \psi^\mu A_\mu$ and which appears in the calculation of $Q^2$, then vanishes algebraically, as before, without further specifying neither $\psi^\mu$ nor $\mathcal{H}$. \\
{\bf Remark:} 
While nilpotency of $Q$ is a natural condition to impose when BRST-quantizing the world line, one may question the rationale for requiring $Q^2=0$ from the space-time point of view, given that there is no gauge symmetry associated to $Q$, i.e. there are no $Q$-exact states at ghost number zero to be decoupled. We return to this question in section \ref{RR-states} and again in section \ref{sigma_model}, where we propose to interpret $q$ alone as BRST operator and $Q$ as a BRST operator on an extended configuration space where the global $q$-symmetry is gauged.

On the other hand, the presence of the extra spinorial generators allows for extra background fields to be considered. In particular, the odd parity of $q$ can be arranged with the degree-shifting operator $\susp$. Furthermore, deforming the supercharge $q$ by background fields will always give rise to a nilpotent BRST differential, since $[q_0+\delta q,\{q_0+\delta q,q_0+\delta q\}]=0$ identically. This feature is an outcome of associativity of the product between $\vartheta,\varepsilon,\lambda$ and $\mathbb{1}$. One such deformation is given by 
\begin{align}
    \delta q=  (\varphi_\alpha\vartheta^\alpha+\tilde\chi^{\dot{\alpha}}\tilde\vartheta_{\dot\alpha}) \susp .
\end{align}
We will not consider this deformation here since it does not preserve $\mathcal{H}$. A deformation preserving $\mathcal{H}$ is given by 
\begin{align}\label{eq:n1os}
    \delta q=  \phi(\vartheta^\alpha\lambda_\alpha\pm\tilde\vartheta_{\dot\alpha}\tilde\lambda^{\dot{\alpha}}) \susp\,,
\end{align}
with BRST closure inducing the non-minimal coupling 
\begin{align}
    H=-2\mathcal{I} \eta^{\mu\nu} p_\mu p_\nu+(2\pm 2)\phi\psi^\mu\susp p_\mu+2\psi^\mu[p_\mu,\delta q] +2\phi^2\mathcal{I}\susp\mathcal{I}\susp\,.
\end{align}
Denoting $q:= q_0 + \delta q$, the remaining commutator $[H,q]$ then again vanishes identically on $\mathcal{H}$. If we choose the "-" sign in \eqref{eq:n1os}, this deformation, being diagonal in the spinor space, can be interpreted as a Yukawa coupling to a Higgs-like field, i.e. adding a mass to the fermion.\footnote{There are other ways to describe mass deformations by adding to $(x^\mu,\psi^\mu)$ a world line SUSY doublet \cite{Carosi:2021wbi}.} 

Another possible deformation preserving $\mathcal{H}$ is
\begin{align}\label{eq:2A_q}
    \delta q= ( A_{\alpha}^{\;\;\beta}\vartheta^\alpha\lambda_\beta+\tilde A_{\dot\alpha}^{\;\;\dot\beta}\tilde\vartheta_{\dot\beta}\tilde\lambda^{\dot{\alpha}}) \susp\,,
\end{align}
with $A_{\alpha}^{\;\;\beta}$ traceless, describing a non-vanishing $2$-form field background. A possible geometric interpretation of this background is a warped compactification with a spin connection whose only non-vanishing component $\omega_{5\alpha}^{\;\;\;\;\beta}=A_{\alpha}^{\;\;\beta}$ is independent on $x^5$. This identification is interesting in view of interpreting $A_{\alpha}^{\;\;\beta}$ as an RR-background field strength arising from dimensional reduction.  More generally, however, $A_{\alpha}^{\;\;\beta}$ and $\tilde A_{\dot\alpha}^{\;\;\dot\beta}$ are independent. The calculation of $\{q,q\}$ then determines the Hamiltonian as 
\begin{align}
    H=&-2\mathcal{I} \eta^{\mu\nu} p_\mu p_\nu+ A_{\mu\nu}A_{\rho\sigma} \left(g^{\mu\rho}g^{\nu\sigma} + \frac{\mathrm{i}}{2}\epsilon^{\mu\nu\rho\sigma}\right) \vartheta\cdot\lambda  \nonumber \\ \, & + 2\mathrm{i}(  \tilde\vartheta\tilde\sigma^\mu\lambda - \vartheta\sigma^\mu\tilde{\lambda}) A_{\mu\nu} p^\nu - \epsilon^{\mu\nu\rho}{}_{\kappa} (\tilde\vartheta\tilde\sigma^\kappa\lambda - \vartheta\sigma^\kappa\tilde{\lambda}) A_{\mu\nu}p_\rho\, \nonumber\\
    \, & + 2\psi^\rho [p_{\rho}, A_{\alpha\beta}]\vartheta^\alpha \lambda^\beta \susp\quad + \quad ( A 
    \;\to
    \;\tilde{A}) \label{p,dq2}.
\end{align}
where $( A \;\to \;\tilde{A})$ means that we add the corresponding terms with $ A$ replaced by $\tilde{A}$. 
Despite the cumbersome expression, nilpotency for this deformation is guaranteed. In fact, it is just a plain consequence of the associativity of the matrix product between the extended Pauli matrices, induced by working in the Hilbert space that is annihilated by the action of $n\geq 2$ operators $\lambda$ or $ \tilde\lambda$.

\subsection{Chiral theory}\label{sec:chiral_theory}
As mentioned above, in contrast to the $\mathcal{N}=2$ world line the space-time interpretation of the world line BRST charge $Q$ is not obvious beyond it reproducing the equations of motion for the Ramond sector which, in turn, are already implied by the world line super charge $q_0$. The latter is not nilpotent but, if we split $q_0$ as follows: 
\begin{align}\label{eq:q_only_R}
    q_0=\mathbf{q}  +\mathbf{\bar q}=\Psi^\mu p_\mu+\bar\Psi^\mu p_\mu=\tilde{\vartheta}_{\dot{\beta}}\tilde{\sigma}^{\mu \; \dot{\beta} \beta} \lambda_{\beta} \susp p_\mu+\vartheta^{\alpha}\sigma^{\mu}_{\alpha \dot{\alpha}}  \tilde{\lambda}^{\dot{\alpha}}\susp p_\mu\,,
\end{align}
 $\mathbf{\bar q}$ and $\mathbf{q}$ are separately nilpotent when acting on the module of ghost-free wavefunctions in $\mathcal{H}$. For concreteness, $\mathbf{q}$ is nilpotent and, acting on the chiral wave function
\begin{align}\label{eq:ferm_L}
    \ket{\Phi^\susp}=\varphi^\susp_\alpha(x) \vartheta^{\alpha} \susp
\end{align}
already produces the Weyl equation $\tilde\partial\!\!\!/\varphi^\susp =0$. We may also consider $\mathbf{q}$ as an operator on the bigger vector space consisting of both chiralities with 
\begin{align}\label{eq:ferm+Big}
    \ket{\Phi^\susp}=\varphi^\susp_\alpha(x) \vartheta^{\alpha} \susp +\tilde\chi^{\susp \,\dot{\alpha}}(x)\tilde\vartheta_{\dot{\alpha}} \susp
\end{align}
Then, $\mathbf{q} \ket{\Phi}$ leaves $\tilde\chi^{\susp \, \dot{\alpha}}(x)$ unconstrained but, provided $\tilde\chi^{\susp\dot{\alpha}}(x)$ is not in the kernel of $p_{\alpha\dot\alpha}$, then $\tilde\chi^{\susp\dot{\alpha}}(x)\tilde\vartheta_{\dot{\alpha}}\susp$ is $\mathbf{q}$-exact\footnote{For $p^2=0$ one shows this in the light cone gauge.}. Thus, the cohomologies of $\mathbf{q}$ and $q_0$ agree and the nilpotence of $\mathbf{q}$ now acquires a space-time interpretation of decoupling the off-shell anti-chiral Ramond sector. 

Like the operator $q_0$ studied in the last subsection, $\mathbf{q}$ admits deformations. First we consider, 
\begin{align}\label{eq:Atdef}
 \delta \mathbf{q}= s\, \tilde{\vartheta}_{\dot{\beta}}\tilde{A}^{\dot{\beta} \beta} \lambda_{\beta}
 \susp\,,
\end{align}
where $s$ is a small parameter. Since $\{\mathbf{q},\delta \mathbf{q}\}$ vanishes on 1-particle states, $\delta \mathbf{q}$ is a good candidate for an interaction vertex in a world-line graph and indeed with the inner product defined in \eqref{hermitian-metric}
\begin{align}\label{eq:wl3}
    (\ket{\Phi},\delta \mathbf{q}\ket{\Phi^\susp})=-s \,\tilde\chi^*_{\dot\beta}\tilde{A}^{\dot{\beta} \beta}\varphi_\beta^\susp
\end{align}
produces the correct amplitude for an interaction of a Weyl fermion with a Yang-Mills potential. Note that, had we used the deformed Hamiltonian $\delta H$ instead of $\delta\mathbf{q}$, inserting $(\mathrm{d} A)_{\dot\alpha\dot\beta}$ instead of $\tilde{A}^{\dot{\beta} \beta}$ in \eqref{eq:wl3}, this  would lead to the wrong amplitude. This gives further support that for the Ramond sector $\mathbf{q}$ rather than $Q$ is the correct choice as a differential. Furthermore, since $\mathbf{q}+\delta \mathbf{q}$ is nilpotent on 1-particle states, the anti-chiral Ramond sector can still be decoupled in a generic background Yang-Mills potential. 

Next we consider the deformation
\begin{align}
 \delta \mathbf{q}=  s\,{\vartheta}^{{\beta}}{F}_{{\beta} \dot\beta} \tilde\lambda^{\dot\beta}
 \susp\,.
\end{align}
Now, $\{\mathbf{q},\delta \mathbf{q}\}$ no longer vanishes on 1-particle states. The reason for this is readily explained. Such a background does not effect the Weyl equation for $\varphi^\susp_\alpha$. On the other hand,
\begin{align}\label{eq:tc+1}
    (\mathbf{q}+\delta \mathbf{q})\tilde\chi^{\susp\dot{\alpha}}(x)\tilde\vartheta_{\dot{\alpha}}\susp= s\, {\vartheta}^{{\beta}}{F}_{{\beta} \dot\beta} \tilde\lambda^{\dot\beta}\susp \tilde\chi^{\susp \, \dot{\alpha}}(x)\tilde\vartheta_{\dot{\alpha}} \susp ={F}_{{\beta} \dot\beta}\tilde\chi^{\susp \,\dot{\beta}}(x){\vartheta}^{{\beta}}\,.
\end{align}
Eqn. \eqref{eq:tc+1} then sets the anti-chiral Ramond sector to zero for generic ${F}_{{\beta} \dot\beta}$ and thus removes the need for nilpotence of  $\mathbf{q}+\delta \mathbf{q}$, since there are no exact states left in the kernel to be decoupled. In the next section we will  propose to interpret this deformation as RR 1-form background field strength, which removes the previously exact anti-chiral Ramond sector  $\tilde\chi^{\susp\dot{\alpha}}(x)\tilde\vartheta_{\dot{\alpha}}\susp$. The world line amplitude of the remaining, massless chiral fermion does not couple to ${F}_{{\beta} \dot\beta}$.  

For 
\begin{align}\label{eq:tFdd}
     \delta \mathbf{q}= s\, {\tilde\vartheta}^{{\dot\beta}}\tilde{F}_{\dot{\beta} \dot\gamma} \tilde\lambda^{\dot\gamma} \susp
\end{align}
we write $\varphi^\susp_{\alpha}=\varphi^\susp_{(0)\alpha}+s\,\varphi^\susp_{(1)\alpha}+\cdots$. The deformation  $\mathbf{q}+\delta \mathbf{q}$ is not nilpotent. Futhermore, closure implies  
\begin{align}
\tilde p^{\dot\alpha \alpha}\varphi^\susp_{(0)\alpha}=0\quad \text{and}\quad \tilde p^{\dot\alpha \alpha}\varphi^\susp_{(1)\alpha}= \tilde{F}^{\dot\alpha}_{\;
\;\dot\beta}\tilde\chi^{\susp\dot\beta}\,,
\end{align}
describing a deformation of $\varphi^\susp$ by $\tilde{\chi}^\susp$. Finally, we consider the deformation
\begin{align}\label{eq:R_F_aa}
 \delta \mathbf{q}= s\, F_{\alpha}^{\;\;\beta}\vartheta^\alpha\lambda_\beta\susp
\end{align}
which is again not nilpotent and, furthermore sets the chiral Ramond sector $\varphi^\susp$ to zero for generic $F_{\alpha}^{\;\;\beta}$, while the anti-chiral Ramond sector is off-shell but not exact. We take this as an indication that such a deformation should be excluded. We will return to this point in the next sections \ref{RR-states} and \ref{sec:Backgd}.

\section{Multiparticle states}
\label{Multiparticle states}
In the previous sections we limited our observations to 1-particle excitations forming a Hilbert space that is invariant under the action of $\psi^\mu$. However, we can also take the tensor product of $1$-particle states, 
not yet $\psi^\mu$ invariant, and later impose this condition on the tensor space, together with the request that the kets must be in ker$(\mathcal{I}-1)$. These requirements define $\mathcal{H}^{\langle 2 \rangle}$ which is again 
a module for the Clifford algebra, where the Gamma matrices act by chirality change and parity shift.

In particular, 2-particle states (either of the same chirality or of opposite chirality) can be constructed in $\mathcal{H}^{\langle2\rangle}$ if we encode in $\varepsilon$ all the ambiguity in the product of $\vartheta$ with $\susp$:
\begin{equation}
    \varepsilon = \susp \vartheta \susp ,
\end{equation}
the two copies of bosonic Weyl spinors being introduced in section \ref{Resolution}. A pair satisfying all the above conditions to belong to $\mathcal{H}^{\langle2\rangle}$ is for instance 
\begin{align}
   \ket{\dot\alpha\dot\beta}= (\tilde\vartheta_{\dot\alpha}  \tilde\varepsilon_{\dot\beta} - \tilde\vartheta_{\dot\alpha} \tilde\vartheta_{\dot\beta} )\begin{pmatrix}\ket{0}\\ \susp \ket{0}\end{pmatrix} \quad\text{and}\quad \ket{e^\beta_{\;\dot\gamma}}= (\vartheta^\beta \tilde\vartheta_{\dot\gamma} -\vartheta^\beta\tilde\varepsilon_{\dot\gamma}) \begin{pmatrix}
   \ket{0} \\ \susp\ket{0}
   \end{pmatrix} \,\label{eq:2theta_0}.
\end{align}
They belong to ker$(\mathcal{I}-1)$ in $\mathcal{H} \otimes \mathcal{H}$ and are mapped to each other by $\psi^\mu$: 
\begin{equation}
    \psi^\mu \ket{\dot\alpha\dot\beta} =  \sigma^\mu_{\gamma \dot\alpha} \vartheta^\gamma \left(\tilde\vartheta_{\dot\beta} - \tilde\varepsilon_{\dot\beta} \right) \begin{pmatrix} \uparrow \ket{0} \\ \ket{0} \end{pmatrix} = \sigma^\mu_{\gamma \dot\alpha} \ket{e^\gamma{}_{\dot\beta}}.
\end{equation}
The first state corresponds to a spacetime scalar and a $2$-form in space-time, whereas the second state corresponds to a spacetime vector. We will consider such states in the following subsection.  
Further elements in $\text{ker}(\mathcal{I}-1)$ are produced by exchanging each $\vartheta$ (and $\varepsilon$) with its chiral counterpart in \eqref{eq:2theta_0}, $\vartheta \leftrightarrow \tilde{\vartheta}$ (and $\varepsilon\leftrightarrow\tilde\varepsilon$): 
\begin{align}
 \ket{\alpha\beta}= (\vartheta^{\alpha} \varepsilon^{\beta} - \vartheta^{\alpha} \vartheta^{\beta} )\begin{pmatrix}
   \ket{0} \\ \susp\ket{0}
   \end{pmatrix} \quad\text{and}\quad \ket{e_{\dot\beta}^{\;\gamma}}= (\tilde\vartheta_{\dot\beta} \vartheta^{\gamma} -\tilde\vartheta_{\dot\beta}\varepsilon^{\gamma} )\begin{pmatrix}
   \ket{0} \\ \susp\ket{0}
   \end{pmatrix} \,\label{eq:2theta_0+op}\,.   
\end{align}
We see that given the above conditions, the set of allowed combinations\footnote{In this counting we do not distinguish $\susp\ket{0}$ and $\ket{0}$ and neither the different spinor indices.} reduces from $4^2$ down to $2\cdot 4$ (action of $\psi^\mu$ close on the set) and then further to $4$ (when asking to be in $\text{ker}(\mathcal{I}-1)$).

The $\psi^\mu$-invariant subspace of ker$(\mathcal{I}-1)$ in the tensor space 
\begin{align}
    T\mathcal{H}= \mathbb{C} + \mathcal{H} + \mathcal{H}\otimes \mathcal{H}+\mathcal{H}\otimes \mathcal{H}\otimes \mathcal{H}+\cdots\,, \label{tensor_sp}
\end{align}
is still infinite dimensional, as one can observe by increasing the number of creation operators. Let us present constructively the $3$-particles states now. As $\psi^\mu$ closes on the subspace generated by $\vartheta$ and $\tilde\vartheta$, but not on $\varepsilon$ or $\tilde\varepsilon$,
\begin{equation}\label{pmatrix}
    \psi^\mu : \vartheta  \begin{pmatrix}
    \ket{0}\\\susp\ket{0} 
    \end{pmatrix} \to \tilde\vartheta \begin{pmatrix}
    \susp\ket{0} \\ \ket{0} 
    \end{pmatrix}, 
\end{equation}
this dictates that the left outermost creation operator must be a $\vartheta$ or a $\tilde\vartheta$. Therefore the number of states in a set closed under the action of $\psi^\mu$, and formed with $3$ creation operators, is $2\cdot4\cdot4$. Now we must impose that they are eigenstates of eigenvalue $1$ for $\mathcal{I}=\vartheta \lambda + \tilde{\vartheta}\tilde{\lambda}$ in $\mathcal{H}\otimes \mathcal{H}\otimes\mathcal{H}$. Let us point out that the condition of having $\vartheta$ or $\tilde\vartheta$ on the left is less restrictive than the latter. The latter condition reduces the number of states by $4$, yielding a total of 28 possibilities.\footnote{Maybe the general rule is $2^{2n-1} - 4$ for $ n >1$.}  In this way one obtains a set of 28 vectors having eigenvalue $1$ for $\mathcal{I}$ and closing under $\psi^\mu$:
\begin{subequations}
\begin{equation}
    \ket{3^{\alpha \beta}{}_{\dot\gamma}} := \vartheta^\alpha \left(\vartheta^\beta \tilde{\vartheta}_{\dot\gamma} -\tilde\vartheta_{\dot\gamma}\vartheta^\beta\right) \begin{pmatrix}
    \ket{0}\\\susp \ket{0}
    \end{pmatrix}, \quad \ket{3_{\dot\gamma}{}^{ \alpha \beta}}, \quad \ket{3^\alpha{}_{\dot\beta \dot\gamma}}, \quad \ket{3_{\dot{\beta}\dot\gamma}{}^{\alpha}}, \quad \ket{3^{\alpha\beta\gamma}}, \quad \ket{3_{\dot\alpha \dot\beta \dot\gamma}},
\end{equation}
\begin{equation}
    \ket{\bar{3}^{\alpha \beta}{}_{\dot\gamma}}:= \vartheta^\alpha\left(\varepsilon^\beta \tilde{\varepsilon}_{\dot\gamma} -\tilde\varepsilon_{\dot\gamma}\varepsilon^\beta\right) \begin{pmatrix}
    \susp \ket{0} \\ \ket{0}
    \end{pmatrix}, \quad \ket{\bar{3}_{\dot\gamma}{}^{\alpha \beta}}, \quad \ket{\bar{3}^{\alpha}{}_{\dot \beta \dot\gamma}}, \quad \ket{\bar{3}_{\dot\beta \dot\gamma}{}^{\alpha}}, \quad \ket{\bar{3}^{\alpha \beta\gamma}}, \quad \ket{\bar{3}_{\dot\alpha \dot\beta \dot\gamma}}.
\end{equation}
\begin{equation*}
    \ket{\varepsilon 3^{\alpha\beta\gamma}} := \vartheta^\alpha(\vartheta^\beta \varepsilon^\gamma -\varepsilon^\gamma\vartheta^\beta) \begin{pmatrix}
    \ket{0} \\ \susp \ket{0}
    \end{pmatrix}, \quad \ket{\varepsilon 3_{\dot\alpha}{}^{\beta\gamma}},\quad \ket{\varepsilon3^{\alpha\beta}{}_{\dot\gamma} }, \quad \ket{\varepsilon 3^{\alpha\;\gamma}_{\; \;\dot\beta} }, \quad \ket{\varepsilon 3_{\dot\alpha \dot\beta}{}^\gamma}
    \end{equation*}
    \begin{equation}
     \ket{\varepsilon 3_{\dot\alpha \; \dot \gamma}^{\; \; \beta}}, \quad \ket{\varepsilon 3_{\dot\alpha\dot\beta\dot\gamma}}, \quad \ket{\varepsilon 3^{\dot\alpha}_{\; \; \dot\beta\dot\gamma}}
\end{equation}
\begin{align}
 \ket{E^{\alpha\beta\gamma}} := \vartheta^\alpha \left( (\vartheta^\beta -\varepsilon^\beta) (\vartheta^\gamma - \varepsilon^\gamma) \right) \begin{pmatrix}
    \ket{0} \\ \susp \ket{0}
    \end{pmatrix}, \quad +\text{1 chirality change (3)},\notag\\ 
    \quad +\text{2 chirality change (3)}, \quad +\text{3 chirality change (1)\,.}\label{eq:cc1}
\end{align}\label{all_3_part}
\end{subequations}

To convince ourselves that the set of states is closed under the action of $\psi^\mu$ let us look for example at $\ket{3^{\alpha\beta}{}_{\dot\gamma}}$:
\begin{align}
    \psi^\mu \ket{3^{\alpha \beta}{}_{\dot\gamma}} = -\tilde{\sigma}^{\mu \, \dot\alpha \alpha} \ket{\bar{3}_{\dot\alpha\dot\gamma}{}^{\beta}}\,.
\end{align}
As a consequence of the above conditions, the commutator of $\mathcal{I}$ with $\psi^\mu$ \eqref{first-gamma} vanishes on the subspace $\mathcal{H}^{\langle3\rangle}$ of $\mathcal{H}\otimes \mathcal{H} \otimes \mathcal{H}$ spanned by these states, 
\begin{equation}
    [\psi^\mu, \mathcal{I}]\vert_{\ket{3}} = 0.
\end{equation}
Upon substituting the explicit expression of  $\psi^\mu$ and $\mathcal{I}$ one finds that this is equivalent to imposing the condition that the action of a pair of annihilators on any 3-spinor $\ket{3}$ gives always zero. This is an alternative way to characterize $n$-particle states.  

We could go on and classify all the $n$-particles states for finite $n$, but this is beyond the scopes of the present article. For future reference, though, let us display one pair of allowed states in $\mathcal{H}^{\langle4\rangle}$:
\begin{equation}
 \vartheta^\alpha (\vartheta^\beta - \varepsilon^\beta)(\varepsilon^\gamma \vartheta^\delta - \vartheta^\delta\varepsilon^\gamma) \begin{pmatrix}
\ket{0}\\ \susp \ket{0}
\end{pmatrix}, \quad \tilde\vartheta^\alpha (\vartheta^\beta - \varepsilon^\beta)(\varepsilon^\gamma \vartheta^\delta - \vartheta^\delta\varepsilon^\gamma) \begin{pmatrix}
\susp \ket{0}\\\ket{0} \end{pmatrix}  .\label{4-part}
\end{equation}
Let us also point out that a wise strategy in order to build higher particles states is to pile up the linear combination $(\vartheta -\varepsilon)$ in front of the ground state, i.e.:
\begin{equation}
    \vartheta^\alpha \left(\vartheta -\varepsilon\right)^\beta \left(\vartheta -\varepsilon\right)^\gamma \cdots \begin{pmatrix} \ket{0} \\ \susp \ket{0}\end{pmatrix}\,.
    \label{multi_part}
\end{equation}

From the above we then conclude that there is an embedding of the $\mathcal{N}=1$ world line in the "tensor" theory defined on a $\psi^\mu$-invariant subspace of $T\mathcal{H}$ in the kernel of $\mathcal{I}-1$ and preserved by $Q_0$. Furthermore, since $Q_0$ preserves the total number of $\vartheta + \tilde\vartheta$ it preserves each term in the sum \eqref{tensor_sp}.

\subsection{RR-states}\label{RR-states}
As seen previously, the 1-particle states reproduce the massless Dirac fermion. In $\mathcal{H}^{\langle2\rangle}$ we may expect RR-states with the two factors in the tensor product representing the left- and right-moving zero mode of the string respectively. Let us reproduce the $2$-particles states \eqref{eq:2theta_0} here for convenience: 
\begin{align}
   \ket{\dot\alpha\dot\beta}= (\tilde\vartheta_{\dot\alpha}  \tilde\varepsilon_{\dot\beta} - \tilde\vartheta_{\dot\alpha} \tilde\vartheta_{\dot\beta} )\begin{pmatrix}\ket{0}\\ \susp \ket{0}\end{pmatrix} \quad\text{and}\quad \ket{e^\beta_{\;\dot\gamma}}= (\vartheta^\beta \tilde\vartheta_{\dot\gamma} -\vartheta^\beta\tilde\varepsilon_{\dot\gamma}) \begin{pmatrix}
   \ket{0} \\ \susp\ket{0}
   \end{pmatrix} \,,\label{2theta_N=1}
\end{align} 
\begin{align}
   \ket{\dot\alpha\dot\beta}= (\tilde\vartheta_{\dot\alpha}  \tilde\vartheta_{\dot\beta} - \tilde\vartheta_{\dot\alpha}\susp \tilde\vartheta_{\dot\beta}\susp )\begin{pmatrix}\ket{0}\\ \susp \ket{0}\end{pmatrix} \quad\text{and}\quad \ket{e^\beta_{\;\dot\gamma}}= (\vartheta^\beta\susp  \tilde\vartheta_{\dot\gamma}\susp -\vartheta^\beta\tilde\vartheta_{\dot\gamma}) \begin{pmatrix}
   \ket{0} \\ \susp\ket{0}
   \end{pmatrix} \,,\label{2theta_N=1}
\end{align} 
with $\psi^\mu$-action,
\begin{align}\label{eq:closure}
   \psi^\mu\ket{\dot\alpha\dot\beta}=(\sigma^\mu)_{\gamma\dot\alpha}\ket{e^\gamma_{\;\dot\beta}}\quad\text{and}\quad\psi^\mu\ket{e^\gamma_{\;\dot\beta}}=(\tilde\sigma^\mu)^{\dot\gamma\gamma}\ket{\dot\gamma\dot\beta}\,.
\end{align}
The reader should bear in mind that $\psi^\mu$ exchanges the ground states $\ket{0}$ and $\susp\ket{0}$, though we will not always display this for notational simplicity. A generic state constructed from \eqref{2theta_N=1} is thus given by\footnote{There is a second copy of RR 1- and 2-forms, obtained by replacing the two states \eqref{2theta_N=1} by \eqref{eq:2theta_0+op}.}
\begin{align}
    \ket{F,\tilde F}=F_\beta^{\;\dot\gamma}(x)\ket{e^\beta_{\;\dot\gamma}}+\tilde F^{\dot\alpha\dot{\beta}}(x)\ket{\dot\alpha\dot\beta}\label{eq:gs_N=1}\,.
\end{align}
Recalling that from \eqref{BRSTN10} and \eqref{first-gamma} 
\begin{align}
    q_0=\psi^\mu p_\mu =\left(\vartheta^{\alpha}\sigma^{\mu}_{\alpha \dot{\alpha}}  \tilde{\lambda}^{\dot{\alpha}} +\tilde{\vartheta}_{\dot{\beta}}\tilde{\sigma}^{\mu \; \dot{\beta} \beta} \lambda_{\beta}\right) \susp p_\mu\,
\end{align}
we have for the vector 
\begin{align}\label{eq:RR_vecN_1}
    \ket{F^{[1]}}&=F_\alpha^{\;\dot\beta}\ket{e^\alpha_{\;\dot\beta}}\,, \quad q_0\ket{F^{[1]}}=(\tilde p)^{\dot\alpha\gamma} F_\gamma^{\;\dot\beta}\ket{\dot\alpha\dot\beta}\,.
\end{align}
Thus $q_0\ket{F^{[1]}}=0$ implies that $\delta F^{[1]}= \mathrm{d}F^{[1]}=0$, for $\delta:= \star \mathrm{d} \star$ the co-differential constructed by means of the Hodge star operator. Thus $Q_0F^{[1]}=0$ implies that $F^{[1]}$ is closed and co-closed which is consistent with interpreting $F^{[1]}$ as the analogue of the RR 1-form field strength in type IIB string theory.\footnote{Note that there is no ghost number -1 state ($\beta\ket{0}=b\ket{0}=0$) since there is no gauge symmetry here in analogy to the Dirac equation for the 1-particle states.} 
For $\ket{\tilde F}=\tilde F^{\dot\alpha\dot{\beta}}\ket{\dot\alpha\dot\beta}$ we get from $q_0\ket{\tilde F}=0$, 
\begin{align}
    \mathrm{d} F^{[0]}+\mathrm{i}\delta F^{[2]}-\frac12*\mathrm{d}F^{[2]}=0 \label{ker q}\,,
\end{align}
which, taking real and imaginary parts, implies $\delta F^{[2]}=0$ and $\mathrm{d} F^{[0]}-\frac12*\mathrm{d}F^{[2]}=0$. This in turn implies  $\delta \mathrm{d} F^{[0]}=  \delta \mathrm{d} F^{[2]}=0$, but falls short of the Bianchi identities $ \mathrm{d} F^{[0]}=0$ and $\mathrm{d} F^{[2]}=0$ needed for an interpretation as RR-field strengths. However, if we set the $0$-form to a constant (as in massless type IIA SUGRA), we recover the necessary Bianchi identity for $F^{[2]}$. Again, the kernel of $q_0$ and $Q_0$ are identical and since there are no ghost number -1 states in the spectrum,  the cohomology of $Q_0$ and $q_0$ in ghost number zero are identical. We will hence set $F^{[0]}$ to a constant in what follows.

To continue  we split $q_0$ as we did for the Ramond sector in section \ref{sec:chiral_theory}.  The general 2-particle state in the kernel of $\mathbf{q}$ is now of the form
\begin{align}
    \ket{F,\tilde F} =F_\alpha^{[1]\dot\beta}\ket{e^\alpha_{\;\dot\beta}}+\tilde A^{[1]\dot\alpha}{}_{\beta}\ket{e_{\dot\alpha}^{\;\beta}} +F^{[2]}_{\;\;\;\alpha\beta}\ket{\alpha\beta}+\tilde F^{[2]\dot\alpha\dot\beta}\ket{\dot\alpha\dot\beta}\,, \label{eq:RR_vecN_1+op_new}
\end{align}
with $F^{[2]}$ and $F^{[1]}$ closed and co-closed, while there is no condition on $\tilde A^{[1]}$ and $\tilde F^{[2]}$. This is still consistent with interpreting $\tilde A^{[1]}$ as an off-shell Maxwell field as in the case for the deformation \eqref{eq:Atdef}. However, if we work with $\Omega^n(M, \mathbb{C})$, $n=0,1,2$, we can solve the conditions for \,$\mathbf{q}$-exactness. More precisely, if $\tilde A$ and $\tilde F$ are not in the kernel of $p_{\alpha \dot\alpha}$, there exists a 
\begin{align}
    \ket{G}=G_\alpha^{[1]\dot\beta}\ket{e^\alpha_{\;\dot\beta}}+G^{[2]}_{\;\;\;\alpha\beta}\ket{\alpha\beta}\,,
    \label{eq:RR_vecN_1+op}
\end{align}
such that \begin{align}
    \tilde A^{[1]\dot\alpha}{}_{\beta}\ket{e_{\dot\alpha}^{\;\beta}}+\tilde F^{[2]\dot\alpha\dot\beta}\ket{\dot\alpha\dot\beta}=\mathbf{q}\ket{G}\,. \label{exact-F_right}
\end{align}
Thus, the cohomology $H_{\mathbf{q}}(\mathcal{H}^{\langle2\rangle},\mathbb{C})$ of complex space-time fields is again identical to that of $q_0$. 

\subsection{Background fields}\label{sec:Backgd}
It turns out that there is a correspondence between RR-states and deformations of $\mathbf{q}$. For instance,  $\ket{F^{[L]}}=F_\alpha^{\;\;\;\dot\beta}\ket{e^\alpha_{\;\dot\beta}}+F_{\alpha\beta}\ket{\alpha\beta}$ can be written as $\ket{F^{[L]}}=\delta\mathbf{q}\ket{\Omega}$, where
\begin{align}\label{eq:d_q_F_F}
\delta\mathbf{q}= F_{\alpha\dot\beta}\vartheta^{\alpha} \tilde{\lambda}^{\dot{\beta}}\susp+ F_{\alpha\beta}\vartheta^{\alpha}\lambda^\beta\susp\,,
\end{align}
and 
\begin{align}
     \ket{\Omega}=\left(\tilde\vartheta_{\dot\gamma}(\tilde\vartheta^{\dot\gamma}-\tilde\varepsilon^{\dot\gamma})\susp+\vartheta^\gamma(\vartheta_\gamma-\varepsilon_\gamma)\susp\right)
\end{align}
is a reference state in $\mathcal{H}^{\langle 2\rangle}$. We can thus interpret $\delta\mathbf{q}$ as the world line\footnote{Strictly speaking we have not defined the world line that gives rise to $\mathbf{q}$ only. We will comment on this in section \ref{sigma_model}.} analog of a RR-vertex operator in string theory with $\mathbf{q}$ closure,  $\{\mathbf{q},\delta\mathbf{q}\}\ket{\Omega}=0$, being equivalent to the linearized equations of motion for the RR-field strengths. Given this identification one may wonder if there is a non-linear generalization 
of $\mathbf{q}+\delta\mathbf{q}$ with nilpotency 
being tantamount to the RR-field equations as was found in \cite{Berkovits:2001ue}, \cite{Adamo:2014wea} and \cite{Dai:2008bh,Bonezzi:2018box,Bonezzi:2020jjq,Grigoriev:2021bes,Carosi:2021wbi} for the massless NS-sector in the pure spinor string, the ambitwistor string and the world line respectively. 

As discussed in section \ref{backgrounds}, $Q^2=0$ is necessary to decouple pure gauge states. In \eqref{eq:RR_vecN_1+op} we decoupled the off-shell part of $\ket{\tilde F}$, using $\mathbf{q}^2=0$ in the absence of background fields. To see if this still holds let us first analyze the kernel of $\mathbf{q}+\delta\mathbf{q}$, with one non-vanishing background field at the time. It is convenient to treat $\delta \mathbf{q}$ as a small perturbation by introducing a formal parameter $s <\!< 1$.

\begin{itemize}
    \item For 
 \begin{align}
     \delta\mathbf{q}= s\,F_{\alpha\dot\beta}\vartheta^{\alpha} \tilde{\lambda}^{\dot{\beta}}\susp,
 \end{align}
we have 
\begin{align}\label{eq:F1_back}
   \delta\mathbf{q}\ket{F}\equiv 0\quad\text{and}\quad \delta\mathbf{q}\ket{\tilde F}=-s\,F_{\alpha\dot\alpha}\left(\tilde F_{\;\;\;\beta}^{\dot\alpha}\ket{\alpha\beta\susp}+\tilde F^{\dot\alpha\dot\beta}  \ket{e^{\alpha}{}_{\dot\beta}\susp } \right)\,.
\end{align}
Thus, the equation for $\ket{F}$ is not affected by an $F_{\alpha\dot\beta}$ background (i.e. it still requires closure and co-closed w.r.t.\,the de Rham differential $\mathrm{d}$) indicative of a free RR 1-form theory, while $\ket{\tilde F}$ must be set to zero in analogy with the R state in \eqref{eq:tc+1}. Furthermore, $\mathbf{q} + \delta \mathbf{q}$ is not nilpotent. This is consistent with the fact that the previously exact terms are no longer in the kernel of the deformed differential  $\mathbf{q} + \delta \mathbf{q}$.
\item For 
\begin{align}
     \delta \mathbf{q}= s\, {\tilde\vartheta}^{{\dot\beta}}\tilde{F}_{\dot{\beta} \dot\gamma} \tilde\lambda^{\dot\gamma} \susp,
\end{align}
we write, in analogy with \eqref{eq:tFdd}, $\ket{F}=\ket{F_{(0)}}+s\, \ket{F_{(1)}}+\cdots$ to find from $\mathbf{q} + \delta \mathbf{q}$-closure at first order in $s$:
\begin{subequations}
\begin{align}
p^{\dot\alpha \alpha}F^{\;\;\;\;\;\;\dot\beta}_{(1)\alpha}&= \tilde{F}^{\dot\alpha}_{\;\;\;
\dot\gamma}    \tilde F^{\dot\gamma\dot\beta}\,, \label{F_1:F2-tilde_defo}\\
p^{\dot\alpha \alpha}F_{(1)\alpha\beta}&= \tilde{F}^{\dot\alpha}_{\;\;\;
\dot\gamma}    \tilde F^{\dot\gamma}_{\;\;\;\beta}\,, \label{F_2:F2-tilde_defo}
\end{align}
\end{subequations}
meaning that for $s>0$, $F^{[1]}=F_{\alpha\dot\beta}$ in \eqref{F_1:F2-tilde_defo} is deformed by a $\tilde{F}_{\dot{\beta} \dot\gamma}$ background, in analogy with the Ramond state of \cref{eq:tFdd}. The r.h.s. can also be viewed as originating  from a Chern-Simons term 
\begin{align}\label{eq:CS}
   \int\limits_{\mathrm{R}^4}C^{[0]}\tilde F\wedge\tilde F\,,\quad F^{[1]}=\mathrm{d} C^{[0]}\,.
\end{align}
The second equation \eqref{F_2:F2-tilde_defo}, with non-vanishing $\tilde F^{\dot\gamma}_{\;\;\;\beta}$ on the r.h.s. enforces a non-vanishing $F_{\alpha\beta}$ background. 
\item On the other hand, for a $F_{\alpha\beta}$ background 
\begin{align}\label{eq:d_q_F+s}
\delta\mathbf{q}=  s\,F_{\alpha\beta}\vartheta^{\alpha}\lambda^\beta\susp\,,
\end{align}
we find that $\ket{F_{(0)}}$ is forced to zero by the closure condition, while $\ket{\tilde F}$ is off-shell and not $\mathbf{q}+\delta \mathbf{q}$-exact, again in analogy with the Ramond sector in \eqref{eq:R_F_aa}. 
\item Finally, the deformation 
\begin{align}
 \delta \mathbf{q}= s\, \tilde{\vartheta}_{\dot{\beta}}\tilde{F}^{\dot{\beta} \beta} \lambda_{\beta}
 \susp
\end{align}
leads to a nilpotent differential on 2-particle states without any conditions on $\tilde{F}^{\dot{\beta} \beta}$. Furthermore, we have 
\begin{align}
   \delta\mathbf{q}\ket{\tilde F}\equiv 0\quad\text{and}\quad \delta\mathbf{q}\ket{F}=-s\,\tilde F^{\dot\alpha\alpha}\left(F_{(0)\alpha}^{\;\;\;\;\;\;\dot\beta}\ket{\dot\alpha\dot\beta\susp}+F_{(0)\alpha\beta}  \ket{e_{\dot\alpha}{}^{\beta}\susp } \right)\,,
\end{align}
or, equivalently, 
\begin{subequations}
\begin{align}
p^{\dot\alpha \alpha}F^{\;\;\;\;\;\;\dot\beta}_{(1)\alpha}&= -\tilde{F}^{\dot\alpha\alpha}   F_{(0) \alpha}{}^{\dot\beta}\,, \label{tF_1:F1_defo}\\
p^{\dot\alpha \alpha}F_{(1)\alpha\beta}&= -\tilde{F}^{\dot\alpha\alpha}    F_{(0)\alpha\beta}\,, \label{tF_2:F2-tilde_defo}
\end{align}
\end{subequations}
in analogy with \eqref{eq:wl3}.
\end{itemize}

In view of this analysis, we come to the conclusion that one consistent on-shell theory for the RR-fields is obtained by setting the $\tilde{F}^{\dot{\beta} \beta}$ and $F_{\alpha\beta}$ backgrounds and states to zero, with  the remaining sector being preserved by \[\mathbf{q}+\delta\mathbf{q} = \tilde\vartheta \slashed{\tilde p} \lambda \susp + s(\tilde\vartheta \tilde F^{[2]} \tilde\lambda + \vartheta F^{[1]} \tilde\lambda) \susp . \] In addition an $F^{[1]}$ background forces the fluctuations of $\tilde F^{[2]}$ to zero by \eqref{eq:F1_back}. Another possibility is to consider the theory of the two 1-form background field strengths\footnote{One may be tempted to identify one of them with the RR 1-form field strength and the other with the Hodge dual of the Kalb-Ramond curvature. }   $F_{\alpha\dot\beta}$ and $\tilde{F}^{\dot{\beta} \beta}$. Then \eqref{eq:F1_back} again implies that a non-vanishing $F_{\alpha\dot\beta}$ obstructs fluctuations of $\tilde{F}^{\dot{\beta} \beta}$ while a non-vanishing $\tilde{F}^{\dot{\beta} \beta}$ background merely  deforms fluctuations of $F_{\alpha\dot\beta}$ by \eqref{tF_1:F1_defo}. 
This highlights the asymmetry between \eqref{eq:F1_back} on one side and \eqref{F_1:F2-tilde_defo} and \eqref{tF_1:F1_defo} on the other: The left spinor index on the r.h.s. of \eqref{F_1:F2-tilde_defo} is a dotted one, which means that it can be compensated by acting with $\mathbf{q}$ on (a deformation of) $F$. On the other hand, the left spinor index of the r.h.s. of \eqref{eq:F1_back} is un-dotted, which means that it cannot be compensated by acting with $\mathbf{q}$ on any state. This appears to be in contradiction with the existence of an inner product compatible with cyclicity, required to build an action, such as the Chern-Simons term \eqref{eq:CS}, from which these two equations are obtained by a variational principle. In the end, if we insist on a action, we seem to be left with just the RR 1-form field strength in the non-linear theory.

We also note that while $F_{\alpha\beta}$ satisfies the correct linear equations for an RR 2-form, it does not seem to be possible to continuously promote it to a background field since it takes the theory off-shell. 

\subsection{A cochain complex for $T\mathcal{H}$}
Let us now shift the focus on the higher particle states of $\mathcal{H}^{\langle n \rangle}$. For this we first want to explicitly point out that any deformed BRST differential $Q$, for an $\mathcal{N}=1$ worldline, is ensured to be nilpotent in $\mathcal{H}^{\langle n \rangle}$ for any $n$. Regarding the chiral theory for $\mathcal{H}^{\langle 2 \rangle}$ and the $\mathbf{q}$-cohomology, the coefficient functions of the states, which are $0,1,2$-forms on spacetime\footnote{Though the $0$-form has been set to a constant.}, are $\tilde{p}^{\dot\alpha \alpha}$ closed or exact. 
The exact ones are obtained by $\tilde{p}^{\dot\alpha \alpha}$ acting on another $0$-, $1$- or $2$-form (left). In particular, they are \emph{not} $p_{\beta \dot\beta}$-closed. Here we are exploring if exact states in $\mathcal{H}^{\langle n \rangle}$ can be viewed as being in the image of a suitable differential action on $\mathcal{H}^{\langle n-1 \rangle}$. 
To this aim, let $\rho_{\alpha}$ and $\tilde\rho^{\dot\alpha}$ be two constant bosonic spinors. Define 
\begin{equation}
    D= d + \tilde{d} := (\varepsilon - \vartheta)^\alpha \rho_{\alpha} + (\tilde\varepsilon - \tilde\vartheta)_{\dot\alpha} \tilde\rho^{\dot\alpha},
\end{equation}
then $\mathcal{H}^{\langle 1 \rangle} \oplus \mathcal{H}^{\langle 2 \rangle} \oplus \mathcal{H}^{\langle 3 \rangle} \oplus \dots$ is a right module for $D$. Furthermore $[D,\psi^\mu]=0$, since one acts from the right and the other from the left. Thus, starting from the subspace $\mathcal{H}$ spanned by \eqref{eq:kets}, $D$ acts by creating a $\psi^\mu$ invariant subspace of $T\mathcal{H}$. In tensor degree $n$, it is the subspace of separable states given by $(n-1)$ tensor products of $\rho$'s, in tensor product with the $n=1$ spinor $\chi$. Since $\rho_\alpha$ (and $\tilde\rho^{\dot\alpha}$) do not depend on $x$, we can form a differential defined by $\mathbf{D}\equiv \mathbf{q}\circ D$, with $\mathbf{q}$ as in \eqref{eq:q_only_R}
\begin{equation}
   \mathcal{H}^{\langle 1 \rangle} \xrightarrow{\mathbf{D}} \mathcal{H}^{\langle 2\rangle}  \xrightarrow{\mathbf{D}}  \mathcal{H}^{\langle 3 \rangle} \xrightarrow{\mathbf{D}} \dots \label{cochain_compl}
\end{equation}
so that $\big(\bigoplus^n_{i=1} \mathcal{H}^{\langle i\rangle}, \mathbf{D}\big)$ acquires the structure of a chain complex. $\mathbf{D}^2 = 0$ follows from the fact that $\mathbf{q}^2 =0$ and that $\mathbf{q}$ and $D$ commute, 
\begin{align}
    \mathbf{q}\circ D=D\circ\mathbf{q}\,.
\end{align}

Given the structure in \eqref{cochain_compl} one can now address the question of the $\mathbf{D}$ cohomology and its relation to the $\mathbf{q}$ cohomology of \cref{RR-states}. We will see that $\mathbf{q}$-closed or exact states are naturally present in the $\mathbf{D}$-cohomology. In $\vartheta$-degree $1$ the cohomology corresponds to the kernel of $\mathbf{D}$:
\begin{align*}
H^1_\mathbf{D}( \mathcal{H}^{\langle 1 \rangle}, \mathbb{C}) = \text{ker}(\mathbf{D})\vert_{\mathcal{H}^{\langle 1 \rangle}} = & \big\{ \varphi_\alpha \ket{\alpha}, \, \tilde\chi^{\dot\alpha}\ket{\dot\alpha} \in \mathcal{H}^{\langle 1 \rangle} \; \text{s.t.}\;  (\tilde p^{\dot\gamma\gamma} \varphi_\gamma) (\rho_\alpha \ket{e_{\dot\gamma}{}^{\alpha}} + \tilde\rho^{\dot\alpha} \ket{\dot\gamma \dot\alpha}) =0 \big\} 
\end{align*}
so we recover once again the Weyl equation for the Weyl spinor $\varphi_\alpha$. For the second cohomology group,
\[ H^2_\mathbf{D}( \mathcal{H}^{\langle 2 \rangle}, \mathbb{C}) =  \ker(\mathbf{D}:\mathcal{H}^{\langle 2 \rangle} \mapsto \mathcal{H}^{\langle 3 \rangle}) / \text{im}(\mathbf{D}: \mathcal{H}^{\langle 1 \rangle} \mapsto \mathcal{H}^{\langle 2 \rangle}). 
\]
Therefore one finds that in this cohomology group there are fields $\tilde F$ (borrowing the notation of \cref{RR-states}), as well as $F$. Being in the kernel implies the condition:
\begin{equation}
    (\tilde{p}^{\dot\gamma\gamma} F^{[2]}_{\gamma\beta})\left( \rho_\alpha \ket{E_{\dot\gamma}{}^{\beta\alpha}} + \tilde\rho^{\dot\alpha} \ket{E_{\dot\gamma \; \dot\alpha}{}^\beta} \right) + (\tilde{p}^{\dot\gamma\gamma} F^{[1]\;\dot\beta}_{\; \alpha}) \left( \rho_\alpha \ket{E_{\dot\gamma\dot\beta}{}^\alpha} + \tilde\rho^{\dot\alpha} \ket{E_{\dot\gamma \dot\beta \dot\alpha}} \right) =0\,.
\end{equation}
In the image one finds $\tilde F^{[2]\dot\alpha\dot\beta} = \tilde p^{\dot\alpha \alpha} \varpi_\alpha(x) \tilde{\rho}^{\dot\beta}$ and $ A^{[1]\dot\alpha}{}_{\beta} = \tilde p^{\dot\alpha\alpha}\varpi_\alpha(x) \rho_\beta$. Thus, contrary to the $\mathbf{q}$-cohomology, now we observe that $\mathbf{D}$-exact 2-particles states have coefficients given by a derivative w.r.t.\,$x$ on the tensor product of spinors, where one of them is constant so that $H^2_{\mathbf{D}}( \mathcal{H}^{\langle 2\rangle})$ contains $H^2_{\mathbf{q}}( \mathcal{H}^{\langle 2\rangle})$ but not vice versa. 

An equivalent way of expressing the quotient is to interpret the differential $\mathbf{D}$ as the generator of a global, odd symmetry by replacing $\rho_\alpha$ and $\tilde\rho^{\dot\alpha}$ by anticommuting parameters $\epsilon_\alpha$ and $\tilde\epsilon^{\dot\alpha}$. Then the
"super fields", 
\begin{subequations}
\begin{align}
    \ket{\phi\susp}&=\phi^\susp_\alpha\vartheta^\alpha\susp \ket{0} +\tilde F^{\dot\alpha}{}_{\beta}\ket{e_{\dot\alpha}^{\;\;\beta}}+\tilde F^{\susp\dot\alpha\dot\beta} \ket{\dot\alpha\dot\beta \susp}\\
    \delta_\epsilon \tilde F^{\dot\alpha}{}_{\beta}&= (p^{\dot\alpha\alpha}\phi^\susp_\alpha)\epsilon_\beta\,, \qquad \qquad \delta_\epsilon \tilde F^{\susp\dot\alpha\dot\beta}=-(p^{\dot\alpha\alpha}\phi^\susp_\alpha)\tilde\epsilon^{\dot\beta}\,, \qquad \qquad \delta_\epsilon \phi^\susp_\beta=0\quad\nonumber\\
\ket{F}=&F_\alpha{}^{\dot\beta}\ket{e^\alpha_{\;\dot\beta}}+G_{\;\;\;\;\;\gamma}^{\dot\alpha\dot\beta}\ket{E_{\dot\alpha\dot\beta}^{\;\;\;\;\;\gamma}}+G_{}^{\dot\alpha\dot\beta\dot\gamma}\ket{E_{\dot\alpha\dot\beta\dot\gamma}}\,,\\
\delta_\epsilon G_{\;\;\;\;\;\;\beta}^{\dot\alpha\dot\beta}& = (p^{\dot\alpha\alpha} F_\alpha{}^{\dot\beta})\epsilon_\beta\,, \qquad \qquad \delta_\epsilon G_{\;\;\;\;\;\;}^{\dot\alpha\dot\beta\dot\gamma} = (p^{\dot\alpha\alpha} F_\alpha{}^{\dot\beta})\tilde\epsilon^{\dot\gamma}\,, \qquad \qquad \delta_\epsilon F_\alpha{}^{\dot\beta}=0\,,\nonumber \\
\ket{\mathrm{F}} =& F_{\alpha \beta}\ket{\alpha\beta} + G^{\dot\alpha}{}_{\beta\gamma} \ket{E_{\dot\alpha}{}^{\beta\gamma}} + G^{\dot\alpha \; \dot\gamma}_{\;\beta} \ket{E_{\dot\alpha \; \dot\gamma}^{\;\beta}} \\
\,  \delta_\epsilon G^{\dot\alpha}{}_{\beta\gamma}& = \tilde p^{\dot\alpha\alpha}F_{\alpha\beta} \epsilon_\gamma, \qquad \qquad \delta_\epsilon G^{\dot\alpha\;\dot\gamma}_{\; \beta} = \tilde p^{\dot\alpha\alpha}F_{\alpha\beta} \tilde\epsilon^{\dot\gamma} , \qquad \qquad \delta_\epsilon F_{\alpha\beta} = 0, \nonumber
\end{align}
\end{subequations}
where $\ket{E_{\dot\alpha\dot\beta\dot\gamma}}$ is given by the last term in \eqref{eq:cc1} (and similarly $\ket{E_{\dot\alpha\dot\beta}{}^\gamma},  \ket{E_{\dot\alpha \; \dot\gamma}^{\;\beta}}, \ket{E_{\dot\alpha}{}^{\beta\gamma}}$ belong to the same family), form multiplets for a nilpotent transformation generated by $\mathbf{D}_{\dot\alpha}$.

\section{Sigma model description}\label{sigma_model}
Recalling the action of the nilpotent operator $\mathbf{q}$ on a generic vector state \eqref{eq:RR_vecN_1+op} we notice that the kernel of  $\mathbf{q}$ is identical to that of 
\begin{align}
    \mathcal{\tilde C}^{\dot\alpha}=\tilde p^{\dot\alpha \alpha}\lambda_\alpha .
\end{align}
A world line sigma model giving rise to the constraints $\mathcal{\tilde C}^{\dot\alpha}$ and  $\mathcal{C}_{\alpha}=p_{\alpha\dot\alpha}\tilde\lambda^{\dot\alpha}$ is readily constructed as 
\begin{align}\label{eq:sig_1}
    S=\int  \left(p_\mu \dot x^\mu + \vartheta^\alpha \dot\lambda_\alpha + \tilde \vartheta_{\dot\alpha}\dot{\tilde\lambda}^{\dot\alpha} +\tilde w_{\dot \alpha}\tilde p^{\dot\alpha\alpha}\lambda_\alpha +  w^\alpha  p_{\alpha \dot\alpha }\tilde \lambda^{\dot\alpha}\right)\,\mathrm{d}\tau \,,
\end{align}
where $w^\alpha$ and $\tilde w_{\dot \alpha}$ are Grassmann even Lagrange multipliers. The constraints $\mathcal{\tilde C}^{\dot\alpha}$ and  $\mathcal{C}_{\alpha}$ generate the symmetry\footnote{ We write $x^\mu = -\frac{1}{2} x^{\dot\alpha \alpha} \sigma^{\mu}_{ \alpha \dot\alpha}$ and $p_\mu\dot x^\mu = -\frac{1}{2} p_{\alpha \dot\alpha} \dot x^{\dot\alpha \alpha} $.}
\begin{subequations}
\begin{align}
    \delta x^{\alpha\dot\alpha}=&2\tilde\mu^{\dot\alpha}\lambda^\alpha+2 \tilde\lambda^{\dot\alpha}\mu^\alpha\\
    \delta \vartheta^{\alpha}=&\tilde\mu_{\dot\alpha}\tilde p^{\dot\alpha\alpha}\\\delta \tilde\vartheta_{\dot\alpha}=&\mu^\alpha p_{\alpha \dot\alpha}\\
    \delta w^\alpha=&\dot\mu^\alpha\\
    \delta \tilde w_{\dot\alpha}=&\dot{\tilde\mu}_{\dot\alpha}\,.
\end{align}
\end{subequations}
We can use this symmetry to gauge fix $w=\tilde w=0$ and thereby introduce the canonical Grassmann odd ghost pairs, $\{\tau^\alpha, \omega_\beta\}=\delta^\alpha_\beta$ and $\{\tilde\tau_{\dot\alpha}, \tilde\omega^{\dot\beta}\}=\delta_{\dot\alpha}^{\dot\beta}$ with action 
\begin{align}\label{eq:sig_2}
     S=\int \left( p_\mu\dot x^\mu + \vartheta^\alpha \dot\lambda_\alpha + \tilde \vartheta_{\dot\alpha}\dot{\tilde\lambda}^{\dot\alpha}+\omega_\beta\dot \tau^\beta+\tilde\omega^{\dot\beta}\dot{\tilde\tau}_{\dot\beta}\right) \mathrm{d}\tau \,,
\end{align}
and BRST operator 
\begin{align}\label{eq:calo}
   \mathcal{O}= \tau^\alpha\mathcal{C}_{\alpha}+ \tilde\tau_{\dot\alpha}\mathcal{\tilde C}^{\dot\alpha}\,,
\end{align}
which now generates the global symmetry transformation 
\begin{subequations}
\begin{align}\label{eq:odgs}
    \delta x^{\dot\alpha\alpha}=&2\tilde\tau^{\dot\alpha}\lambda^\alpha+2 \tilde\lambda^{\dot\alpha}\tau^\alpha\\
    \delta \vartheta^{\alpha}=&\tilde\tau_{\dot\alpha}\tilde p^{\dot\alpha\alpha}\\
    \delta \tilde\vartheta_{\dot\alpha}=&\tau^\alpha p_{\alpha \dot\alpha}\\
    \delta \omega_\alpha=&p_{\alpha \dot\alpha }\tilde \lambda^{\dot\alpha} \label{eq:odgs2}\\
    \delta \tilde \omega^{\dot\alpha}=&\tilde p^{\dot\alpha\alpha}\lambda_\alpha \,. \label{eq:odgs3}
\end{align}
\end{subequations}
\subsection{Representation on space-time fields}\label{sec:sig_rep}
It is consistent to consider only the chiral part  $\mathcal{O}^{(L)}=\tilde\tau_{\dot\alpha}\mathcal{\tilde C}^{\dot\alpha}$ (or anti-chiral part  $\mathcal{O}^{(R)}=\tau^\alpha\mathcal{C}_{\alpha}$) of the constraint and the corresponding BRST operator on \eqref{eq:calo}. This corresponds to introducing only one auxiliary field (say $\tilde w^{\dot\alpha}$) in \eqref{eq:sig_1} and, accordingly, only one ghost pair $(\tilde\tau^{\dot\alpha}, \tilde\omega_{\dot\beta})$ in \eqref{eq:sig_2}. 
To continue, we choose the representation space to be spanned by (wave) functions
\begin{align}\label{eq:sigma_wave}
    \psi(x,\vartheta^{\alpha},\tilde\tau_{\dot\alpha})\,,
\end{align}
with constraint equation $\mathcal{O}\ket{\psi}=0$. For instance, for 
\begin{align}\label{eq:sigma_phi}
    \ket{\phi}=\phi_\alpha\vartheta^\alpha\,;\quad \mathcal{O}\ket{\phi}=0\,,
\end{align}
we recover the Dirac equation $(p\!\!/\phi)^{\dot\alpha}=0$. The vector 2-particle state
\begin{align}\label{eq:sigma_vec}
    \ket{F}= F_\alpha^{\;\;\dot\beta}\vartheta^{\alpha}\tilde\tau_{\dot\beta}
\end{align}
leads to the equation
\begin{align}\label{eq:caloF}
    0=\mathcal{O}^{(L)}\ket{F}=(p^{\dot\alpha\alpha}F_\alpha^{\;\;\dot\beta})\tilde\tau_{\dot\alpha}\tilde\tau_{\dot\beta}=\tilde\tau^2\,\delta F
\end{align}
which misses the exactness condition $\mathrm{d}F=0$. The latter can, however, be adjusted for by adding an exact contribution 
\begin{align}\label{eq:RR_ghost}
    \mathcal{O}^{(L)} F_{\alpha\beta}\;\vartheta^{\alpha}\vartheta^{\beta}=\tilde p^{\dot\alpha\alpha}F_{\alpha\beta}\vartheta^{\beta}\tilde\tau_{\dot\alpha}\,.
\end{align}
Thus, the cohomology of $\mathcal{O}^{(L)}$ in ghost degree $1$ is again given by a RR 1-form field strength.\footnote{Alternatively, the correct equations for the RR 1-form can be obtained imposing  $
     \mathcal{O}^{(L)}F_\alpha^{\;\;\dot\beta}\vartheta^{\alpha}\tilde\vartheta_{\dot\beta}=0
 $. 
Thus the cohomology at ghost number $0$ in $\{\psi(x,\vartheta^{\alpha},\tilde\vartheta_{\dot\alpha})\}$ and at ghost number $1$ in $\{\psi(x,\vartheta^{\alpha},\tilde\tau_{\dot\alpha})\}$ agree.}

The only non-trivial deformation of $\mathcal{O}^{(L)}$ that preserves the space of functions \eqref{eq:sigma_wave} is 
\begin{align}\label{eq:C-def_vec}
    \delta \mathcal{O}^{(L)}=A^{\alpha\dot\beta}\tilde\tau_{\dot\beta}\lambda_\alpha\,,
\end{align}
which we had identified in the last sections with a Yang-Mills vector potential. Concerning nilpotency (i.e. the decoupling of exact states) we note that 
 \begin{align}
     \{\mathcal{O}^{(L)}+\delta \mathcal{O}^{(L)},\mathcal{O}^{(L)}+\delta \mathcal{O}^{(L)}\}=2\tilde{\tau}_{\dot{\beta}}\tilde{\tau}_{{\dot\alpha}} \lambda_{\beta} \lambda_{\alpha}[{p}^{\beta\dot\beta},  A^{\alpha\dot{\alpha}}]\,,
 \end{align}
vanishes on 1-particle states (Ramond sector) but fails to vanish on the RR space-time ghost \eqref{eq:RR_ghost}. We thus (consistently) exclude this background in the RR sector. A deformation by a RR 1-form background as in section \ref{sec:Backgd} does not appear to arise naturally in this description.

\subsection{Relation to the RNS formulation:}\label{sec:R_RNS}
In order to make contact with the resolved RNS formulation in section \ref{Multiparticle states}, we may consider the action \eqref{eq:sig_2} as the starting point rather than a gauge fixed BRST action together with the constraint $\mathcal{O}$.\footnote{The global symmetry \eqref{eq:odgs} can be promoted to a local one by adding a term $\int\chi \mathcal{O} $ to the world line action \eqref{eq:sig_2} where $\chi$ is fermionic Lagrange multiplier that will give rise to a bosonic ghost $\gamma$. } We then add the extra Lagrange multiplier 
\begin{align}
    \Delta S =\int \tilde\rho_{\dot\alpha}(\tilde\omega^{\dot\alpha}-\tilde\lambda^{\dot\alpha}\susp)+\rho^{\alpha}(\omega_{\alpha}-\lambda_{\alpha}\susp)\equiv\int \tilde\rho_{\dot\alpha}\tilde\Theta^{\dot\alpha}+\rho^{\alpha}\Theta_{\alpha}\,
\end{align}
giving rise to the secondary constraint
\begin{align}
    \{\mathcal{O},\tilde\Theta^{\dot\alpha}\}=\tilde{ \mathcal{C}}^{\dot\alpha}\,,
\end{align}
which commutes with the other constraints. Similarly for $\Theta^\alpha$. Here we assumed that the Poisson bracket between $\susp$ and $\tau^\alpha$ vanishes. 

We can solve the constraint $\tilde\Theta^{\dot\alpha}$ by replacing $\tilde\omega^{\dot\alpha}$ in \eqref{eq:sig_2} by $\tilde\lambda^{\dot\alpha}\susp$.  Then, the variable conjugate to $\tilde\lambda$ is given by
\begin{align}\label{eq:hat_theta}
    \tilde\vartheta_{\dot\alpha}+\tilde\tau_{\dot\alpha}\susp\,,
\end{align}
while $\tilde\vartheta_{\dot\alpha}-\tilde\tau_{\dot\alpha}\susp$ is a cyclic variable and can thus be removed by adding one more Lagrange multiplier without generating new secondary constraints. From the point of view of the BRST action \eqref{eq:sig_2} these linear combinations have inhomogenous ghost number. However, in this subsection the ghosts are interpreted merely as  fermionic world line fields and $\mathcal{O}$ generates the odd global symmetry \eqref{eq:odgs}.

To continue, we impose $\tilde\Theta_{\dot\alpha}$ and $\Theta^\alpha$ on the representation space. In the chiral Ramond sector (at ghost number zero) this has no effect while for the vector \eqref{eq:sigma_vec}, using \eqref{eq:hat_theta}, we replace 
$\tilde\tau_{\dot\alpha}\susp$ with $\tilde\vartheta_{\dot\alpha}$. More precisely, 
\begin{align}\label{eq:F_sig}
    \ket{F}= F_\alpha^{\;\;\dot\beta} \vartheta^\alpha(\tilde\vartheta_{\dot\beta}-\tilde\varepsilon_{\dot\beta})=F_\alpha^{\;\dot\beta}\ket{e^\alpha_{\;\dot\beta}}
\end{align}
solves the constraint $\tilde\Theta_{\dot\alpha}$. Furthermore, $\mathcal{O}^{(L)}$ reproduces $\mathbf{q}$ of section \ref{sec:chiral_theory} after replacing $\tilde\tau_{\dot\alpha}$ with $\tilde\vartheta_{\dot\alpha}\susp$.

\subsection{Relation to the Brink-Schwarz particle}\label{sec:BS_particle}
In analogy with \cite{Berkovits:1993xq} the action \eqref{eq:sig_2} together with the BRST operator \eqref{eq:calo} allows for an alternative interpretation where $\lambda_\alpha, \tilde\lambda^{\dot\alpha}$ and $\vartheta_\alpha,\tilde\vartheta^{\dot\beta}$ are interpreted as ghosts and anti-ghost respectively, while  $(\omega_\alpha,\tau^\alpha,\tilde\tau_{\dot\alpha},\tilde\omega^{\dot\beta})$ are world-line fermions. Then, after the change of variables\footnote{Note, however,  that this transformation is singular at $p_{\alpha\dot\alpha}=0$.}
\begin{align}
   \omega_\alpha=p_{\alpha\dot\alpha} {\tilde\theta}^{\dot\alpha}\,\quad\text{and}\quad \tilde\omega^{\dot\alpha}=\tilde p^{\dot\alpha\alpha} {\theta}_{\alpha}
\end{align}
the transformations \eqref{eq:odgs}, \eqref{eq:odgs2} and \eqref{eq:odgs3} have the interpretation of the BRST-transfor-\\mations originating from the local fermionic transformation
\begin{subequations}
\begin{align}\label{eq:odgs_SUSY}
    \delta x^{\dot\alpha\alpha}=&2\tilde\tau^{\dot\alpha}\epsilon^\alpha+2 \tilde\epsilon^{\dot\alpha}\tau^\alpha\\
    \delta{\tilde\theta}^{\dot\alpha}=&\tilde \epsilon^{\dot\alpha}\\
    \delta{\theta}_{\alpha}=&\epsilon_\alpha\,,
\end{align}
\end{subequations}
while the action \eqref{eq:sig_2} takes the form\footnote{Here we removed the "new" ghost $\lambda$ and $\vartheta$. } 
\begin{align}\label{eq:sig_BS}
     S=\int p_{\alpha\dot\alpha}( -\frac{1}{2}\dot x^{\dot\alpha\alpha} +{\tilde\theta}^{\dot\alpha}\dot\tau^\alpha   + \dot{\tilde\tau}^{\dot\alpha}  {\theta}^{\alpha})   \mathrm{d} t\,.
\end{align}
This bears some similarities with the Brink-Schwarz particle \cite{Brink:1981nb}. In particular, the transformation \eqref{eq:odgs_SUSY} generated by 
\begin{align}\label{eq:S_BS}
    \tilde d_{\dot\alpha}=\tilde p_{\dot\alpha\alpha}\tau^\alpha\,\quad\text{and}\quad  d^{\alpha}= p^{\alpha\dot\alpha}\tilde\tau_{\dot\alpha}\,.
\end{align}
 is a {\it local} fermionic symmetry which, with the replacement $\epsilon_\alpha=p_{\alpha\dot\beta}\kappa^{\dot\beta}$, can be recast into 
 $\kappa$-symmetry. On the other hand, $\tau^\alpha$ and $\tilde\tau_{\dot\alpha}$ are invariant. Also, due to the absence of a $\dot\theta^\alpha$ term, \eqref{eq:sig_BS} does not feature a space-time supersymmetry.

\section{Summary and discussion}
In this article, we explored a way to obtain \emph{spin fields} from a resolution of the RNS world line for the relativistic massless particle. To this aim, we replaced the graded Lie algebra of the world line fields by an associative, but non-Lie, operator algebra \eqref{FFCR0}, \eqref{FFCR1} acting on some representation space of 1, 2 (and in fact also higher) particle states corresponding to Ramond, i.e. space-time fermions, and RR fields respectively, as described in \cref{sec:N=1_BRST} and \cref{Multiparticle states}. Each of these complex vector spaces of $n$-particles defines a sub-complex of the initial BRST complex of the world line theory. 
For $n=1,2$ it describes massless fermions (\cref{sec:N=1_BRST}) and an on-shell RR 1-form and 2-form field strength (\cref{RR-states}) respectively, thus mimicking the RR spectrum of the type II string in four dimensions. An instrumental role in our approach is played by the chiral part $\mathbf{q}$ of the world line supercharge which acts as the differential in the relevant subcomplex (without BRST ghosts) and whose nilpotence has the simple interpretation of decoupling off-shell states. This is in contrast to the BRST differential $Q$ whose nilpotence had no significance for physical states. 

An appealing feature of the present resolution in the chiral theory is that it allows to couple the fields in the representation space to background fields. In particular, we found a correspondence between deformations of the differential by background fields and the RR-states in the representation space. At the non-linear level, the deformation by a RR 1-form field strength removes half of the perturbative states from the spectrum.  Other background fields can be turned on as well. Some of them lead to an off-shell theory while others lead to Chern-Simons terms combining the RR 0-form potential with a 2-form field strength. However, the full set of coupled equations obtained in this way, does not seem to derive from an action, at least not one that involves only 0, 1 and 2-form fields. It would be interesting to understand this better in view of extracting an action for this dynamical system. Nevertheless a consistent truncation involving only the RR 1-form exists as in analogy with the GSO projection in string theory.   

We then constructed a simple world line sigma model for the spin field which reproduces the resolved RNS theory after imposing suitable constraints on the ghosts. We checked that the spinor - and vector cohomologies of the chiral BRST operator for the sigma model agree with the $\mathbf{q}$-cohomology. However, this sigma model does not seem to be related to the Brink-Schwarz particle which was the starting point from the pure-spinor approach \cite{Berger:2002wz}, see \cref{sec:BS_particle}. It would be interesting to better understand the precise relation between the latter and our resolution. For this one should probably adapt our construction to a 10-dimensional target space. A central role in our ansatz is played by the shift operator $\susp$, which does not appear in other constructions. Comparing this with the field redefinition proposed in  \cite{Berkovits:2001us} to relate the pure spinor and RNS formulation one may speculate that $\susp$ can be combined with the super ghost $\gamma$ (or $\delta(\gamma)$) which were not present in our work.

We did not explore possible extensions to massless higher spin degrees of freedom as multi-particle states on the world line, as done for example in \cite{Plyushchay:2003gv}. It would be intriguing to see whether the 3-particles and higher states of \eqref{all_3_part}, \eqref{4-part} and \eqref{multi_part} can have such interpretation. 

\section*{Acknowledgements} We would like to thank N. Berkovits for a correspondence. E.B. acknowledges GA\v{C}R grant EXPRO 19-28628X for financial support. I.S. acknowledges the Excellence Cluster Origins of the DFG under Germany’s Excellence Strategy EXC-2094 390783311 for financial support.\\

\section*{A: Conventions}
{\bf Spinor representations:}\\
Our choice of conventions is based on the invariant tensor
\begin{equation}
\epsilon^{\alpha\beta}\equiv (\mathrm{i}\sigma_2)^{\alpha\beta}=\begin{pmatrix}0&1\cr  -1&0\end{pmatrix}\equiv \epsilon^{\dot\alpha\dot\beta}
\end{equation}
or
\begin{equation}
1=\epsilon^{12}=-\epsilon^{21}=1=-\epsilon_{12}=\epsilon_{21}.
\end{equation}
In particular, $\epsilon^{\alpha \beta} \epsilon_{\alpha \beta} = -2$.

For $Q$ an element of $V\otimes \dot V$, we write 
\begin{equation}\label{Q2}
Q=Q^{\alpha\dot\beta}=q^\mu(\sigma_\mu)^{\alpha\dot\beta}
\end{equation}
Under parity transformation,
\begin{equation}
P:\qquad q=(q^0,\vec{q})\quad\to \quad {}^Pq=(q^0,-\vec{q})
\end{equation}
the bi-spinor $Q$ transforms as 
\begin{equation}
 {}^P\!\!Q=q^0-\vec{q}\cdot\vec{\sigma}=\epsilon^{-1} Q\epsilon\equiv q^\mu(\tilde\sigma_\mu)
\end{equation}
We then have 
\begin{align}
    \tilde\sigma_\mu^{\dot\alpha\alpha}&= \epsilon^{\dot\alpha\dot\beta}\epsilon^{\alpha\beta}(\sigma_\mu)_{\beta\dot\beta}, \label{sigma_tilde}\\
     (\sigma_\mu)_{\alpha\dot\alpha}&= \epsilon_{\alpha\beta}\epsilon_{\dot\alpha\dot\beta}(\tilde\sigma_\mu)^{\dot\beta\beta}.
\end{align}
{\bf Minkowski metric:}
\begin{equation}
    \eta_{\mu \nu} \, \mathrm{d}x^{\mu} \vee \mathrm{d}x^{\nu} = - \mathrm{d}x^{0} \vee \mathrm{d} x^{0} + \delta_{ij} \mathrm{d} x^{i} \vee \mathrm{d}x^{j}. \label{Mink}
\end{equation}
{\bf Fierz identities:}
\begin{align}\label{F1}
    (\sigma^\mu)_{\alpha\dot\alpha}(\sigma_\mu)_{\beta\dot\beta} &= -2\epsilon_{\alpha\beta}\epsilon_{\dot\alpha\dot\beta},\\
     (\tilde\sigma^\mu)^{\dot\alpha\alpha}(\tilde\sigma_\mu)^{\dot\beta\beta} &=-2\epsilon^{\alpha\beta}\epsilon^{\dot\alpha\dot\beta}, \label{F2}\\
     (\sigma^\mu)_{\alpha\dot\alpha}(\tilde\sigma_\mu)^{\dot\beta\beta} &=-2\delta^{\dot\beta}_{\dot\alpha}\delta^\beta_\alpha. \label{F3}\\
     \delta_\beta^\alpha(\sigma^\mu)_{\alpha\dot\alpha}(\tilde\sigma^\nu)^{\dot\beta\beta} &=-\eta^{\mu\nu}\delta^{\dot\beta}_{\dot\alpha}+[\sigma^{\mu\nu}]^{\dot\beta}{}_{\dot\alpha}. 
\end{align}
{\bf Lorentz group $SO(1,3)$ generators:}

\begin{equation}
    \sigma^{\mu \nu} := \frac{\mathrm{i}}{4} \left(\sigma^{\mu} \tilde{\sigma}^{\nu} - \sigma^{\nu} \tilde{\sigma}^{\mu} \right), \qquad \tilde{\sigma}^{\mu \nu} := \frac{\mathrm{i}}{4} \left( \tilde{\sigma}^{\mu} \sigma^{\nu} - \tilde{\sigma}^{\nu} \sigma^{\mu}  \right).
\end{equation}
{\bf Identity involving three generalized Pauli matrices:}
\begin{equation}
    \tilde{\sigma}^{\rho} \sigma^{\mu \nu} = \frac{\mathrm{i}}{2} (g^{\rho \nu} \tilde{\sigma}^{\mu} - g^{\rho \mu}\tilde{\sigma}^{\nu}) - \frac{1}{2} \epsilon^{\rho \mu \nu \delta}\tilde{\sigma}_{\delta}.
    \label{tripl}
\end{equation}
{\bf$SO(1,3)$ algebra:}
\begin{align} \tilde{\sigma}^{\mu \nu} \tilde{\sigma}^{\rho \sigma} &= \frac{1}{4} \left(g^{\mu \sigma} g^{\rho\nu} -g^{\mu\rho}g^{\nu\sigma}\right) + \frac{\mathrm{i}}{4} \epsilon^{\mu\nu\rho\sigma} + \frac{\mathrm{i}}{4} \left(g^{\sigma\nu}\tilde{\sigma}^{\mu\rho} - g^{\nu\rho} \tilde{\sigma}^{\mu\sigma} +g^{\mu\rho}\tilde{\sigma}^{\nu\sigma} - g^{\mu\sigma}\tilde{\sigma}^{\nu\rho}\right), \label{square} \\
[\tilde{\sigma}^{\mu \nu}, \tilde{\sigma}^{\rho \delta}] & =  \frac{\mathrm{i}}{2}\left(g^{\rho\mu}\tilde{\sigma}^{\nu\delta} - g^{\rho\nu} \tilde{\sigma}^{\mu \delta} - g^{\mu\delta} \tilde{\sigma}^{\nu \rho} + g^{\nu\delta} \tilde{\sigma}^{\mu\rho}\right)\\
\epsilon^{\mu\nu}{}_{\rho\kappa} \sigma^{\rho\kappa} &= \mathrm{i} \sigma^{\mu\nu}. \label{observ} \end{align}
{\bf 4-component spinor:}
\begin{equation}\label{cliffr}
\psi = \begin{pmatrix}u\cr 
\tilde v
\end{pmatrix}\,,\quad
\gamma_\mu = \begin{pmatrix}0&\sigma_\mu\cr 
\tilde\sigma_\mu&0
\end{pmatrix}.
\end{equation}
{\bf Charge conjugation:}
\begin{align}
    \psi^c=\eta_c C \gamma^{0T}\psi^*\;,\qquad C=\begin{pmatrix}
    -\mathrm{i}\sigma_2&0\cr 0&\mathrm{i}\sigma_2
    \end{pmatrix}\qquad \eta_c \in \mathbb{C}, \, \vert\eta_c\vert = 1 .
\end{align}

\bibliographystyle{unsrt}
\bibliography{Bibmaster.bib}

\begin{thebibliography}{10}

\bibitem{Friedan:1985ge}
Daniel Friedan, Emil~J. Martinec, and Stephen~H. Shenker.
\newblock {Conformal Invariance, Supersymmetry and String Theory}.
\newblock {\em Nucl. Phys. B}, 271:93--165, 1986.

\bibitem{Green:1983wt}
Michael~B. Green and John~H. Schwarz.
\newblock {Covariant Description of Superstrings}.
\newblock {\em Phys. Lett. B}, 136:367--370, 1984.

\bibitem{Berkovits:2002zk}
Nathan Berkovits.
\newblock {ICTP lectures on covariant quantization of the superstring}.
\newblock {\em ICTP Lect. Notes Ser.}, 13:57--107, 2003.

\bibitem{Berkovits:1995ab}
Nathan Berkovits.
\newblock {SuperPoincare invariant superstring field theory}.
\newblock {\em Nucl. Phys. B}, 450:90--102, 1995.
\newblock [Erratum: Nucl.Phys.B 459, 439--451 (1996)].

\bibitem{Sen:2014dqa}
Ashoke Sen.
\newblock {Gauge Invariant 1PI Effective Action for Superstring Field Theory}.
\newblock {\em JHEP}, 06:022, 2015.

\bibitem{Erler:2015lya}
Theodore Erler, Sebastian Konopka, and Ivo Sachs.
\newblock {Ramond Equations of Motion in Superstring Field Theory}.
\newblock {\em JHEP}, 11:199, 2015.

\bibitem{Pius:2014iaa}
Roji Pius, Arnab Rudra, and Ashoke Sen.
\newblock {Mass Renormalization in String Theory: General States}.
\newblock {\em JHEP}, 07:062, 2014.

\bibitem{Mattiello:2019gxc}
Luca Mattiello and Ivo Sachs.
\newblock {On Finite-Size D-Branes in Superstring Theory}.
\newblock {\em JHEP}, 11:118, 2019.

\bibitem{Maccaferri:2019ogq}
Carlo Maccaferri and Alberto Merlano.
\newblock {Localization of effective actions in open superstring field theory:
  small Hilbert space}.
\newblock {\em JHEP}, 06:101, 2019.

\bibitem{Vosmera:2019mzw}
Jakub Vo\v{s}mera.
\newblock {Generalized ADHM equations from marginal deformations in open
  superstring field theory}.
\newblock {\em JHEP}, 12:118, 2019.

\bibitem{Cho:2018nfn}
Minjae Cho, Scott Collier, and Xi~Yin.
\newblock {Strings in Ramond-Ramond Backgrounds from the Neveu-Schwarz-Ramond
  Formalism}.
\newblock {\em JHEP}, 12:123, 2020.

\bibitem{Sen:2020cef}
Ashoke Sen.
\newblock {D-instanton Perturbation Theory}.
\newblock {\em JHEP}, 08:075, 2020.

\bibitem{Brink:1981nb}
Lars Brink and John~H. Schwarz.
\newblock {Quantum Superspace}.
\newblock {\em Phys. Lett. B}, 100:310--312, 1981.

\bibitem{Berkovits:2001ue}
Nathan Berkovits and Paul~S. Howe.
\newblock {Ten-dimensional supergravity constraints from the pure spinor
  formalism for the superstring}.
\newblock {\em Nucl. Phys. B}, 635:75--105, 2002.

\bibitem{Dai:2008bh}
Peng Dai, Yu-tin Huang, and Warren Siegel.
\newblock {Worldgraph Approach to Yang-Mills Amplitudes from N=2 Spinning
  Particle}.
\newblock {\em JHEP}, 10:027, 2008.

\bibitem{Bonezzi:2018box}
Roberto Bonezzi, Adiel Meyer, and Ivo Sachs.
\newblock {Einstein gravity from the $ \mathcal{N}=4 $ spinning particle}.
\newblock {\em JHEP}, 10:025, 2018.

\bibitem{Bonezzi:2020jjq}
Roberto Bonezzi, Adiel Meyer, and Ivo Sachs.
\newblock {A Worldline Theory for Supergravity}.
\newblock {\em JHEP}, 06:103, 2020.

\bibitem{Bockisch:2022eas}
Daniel Bockisch and Ivo Sachs.
\newblock {Dark Energy and the Spinning Superparticle}.
\newblock 3 2022, arXiv:2203.06014.

\bibitem{Sorokin:1988nj}
Dmitri~P. Sorokin, Vladimir~I. Tkach, Dimitrij~V. Volkov, and Aleksandr~A.
  Zheltukhin.
\newblock {From the Superparticle Siegel Symmetry to the Spinning Particle
  Proper Time Supersymmetry}.
\newblock {\em Phys. Lett. B}, 216:302--306, 1989.

\bibitem{Volkov:1988vf}
Dimitrij~V. Volkov and Aleksandr~A. Zheltukhin.
\newblock {Extension of the Penrose Representation and Its Use to Describe
  Supersymmetric Models}.
\newblock {\em JETP Lett.}, 48:63--66, 1988.

\bibitem{Brink:1976sc}
Lars Brink, Paolo Di~Vecchia, and Paul~S. Howe.
\newblock {A Locally Supersymmetric and Reparametrization Invariant Action for
  the Spinning String}.
\newblock {\em Phys. Lett. B}, 65:471--474, 1976.

\bibitem{Sorokin:1988jor}
Dmitri~P. Sorokin, Vladimir~I. Tkach, and Dimitrij~V. Volkov.
\newblock {Superparticles, Twistors and Siegel Symmetry}.
\newblock {\em Mod. Phys. Lett. A}, 4:901--908, 1989.

\bibitem{Schubert:2001he}
Christian Schubert.
\newblock {Perturbative quantum field theory in the string inspired formalism}.
\newblock {\em Phys. Rept.}, 355:73--234, 2001.

\bibitem{Carosi:2021wbi}
Matthias Carosi and Ivo Sachs.
\newblock {Proca theory from the spinning worldline}.
\newblock {\em JHEP}, 01:135, 2022.

\bibitem{Adamo:2014wea}
Tim Adamo, Eduardo Casali, and David Skinner.
\newblock {A Worldsheet Theory for Supergravity}.
\newblock {\em JHEP}, 02:116, 2015.

\bibitem{Grigoriev:2021bes}
Maxim Grigoriev, Adiel Meyer, and Ivo Sachs.
\newblock {A toy model for background independent string field theory}.
\newblock 6 2021, arXiv:2106.07966.

\bibitem{Berkovits:1993xq}
Nathan Berkovits and Cumrun Vafa.
\newblock {On the Uniqueness of string theory}.
\newblock {\em Mod. Phys. Lett. A}, 9:653--664, 1994.

\bibitem{Berger:2002wz}
Roland Berger, Michel Dubois-Violette, and Marc Wambst.
\newblock {H}omogeneous algebras.
\newblock 2002, arXiv:math/0203035.

\bibitem{Berkovits:2001us}
Nathan Berkovits.
\newblock {Relating the RNS and pure spinor formalisms for the superstring}.
\newblock {\em JHEP}, 08:026, 2001.

\bibitem{Plyushchay:2003gv}
Mikhail Plyushchay, Dmitri Sorokin, and Mirian Tsulaia.
\newblock {Higher spins from tensorial charges and OSp(N|2n) symmetry}.
\newblock {\em JHEP}, 04:013, 2003.

\end{thebibliography}

\end{document}